\documentclass[12pt]{amsart}
\reversemarginpar
\newcommand{\commentout}[1]{}
\usepackage{amsmath}
\usepackage{amssymb}
\usepackage{graphicx}
\usepackage{graphicx}
\usepackage{dcolumn}
\usepackage{amsfonts}
\usepackage{latexsym}

\newcommand{\nwc}{\newcommand}

\newcommand{\lt}{\left}
\nwc{\partz}{\frac{\partial }{\partial z}}
\newcommand{\rt}{\right}
\nwc{\ytil}{\tilde{\by}}
\nwc{\al}{\alpha}
\nwc{\half}{\frac{1}{2}}

\newcommand{\bx}{\mathbf x}

\newcommand{\bp}{\mathbf p}
\newcommand{\br}{\mathbf r}

\newcommand{\by}{\mathbf y}

\nwc{\ztil}{{\tilde z}}
\nwc{\betatil}{{\tilde \beta}}
\nwc{\bC}{{\mathbf C}}
\nwc{\bM}{{\mathbf M}}
\nwc{\bs}{\mathbf s}
\nwc{\nwt}{\newtheorem}

\nwt{proposition}{Proposition}
\nwt{lemma}{Lemma}
\nwt{theorem}{Theorem}

\nwt{remark}{Remark}
\nwt{assumption}{Assumption}
\nwt{definition}{Definition} 

\nwc{\bal}{\begin{align}}
\nwc{\be}{\begin{equation}}
\nwc{\ben}{\begin{equation*}}
\nwc{\bea}{\begin{eqnarray}}
\nwc{\beq}{\begin{eqnarray}}
\nwc{\bean}{\begin{eqnarray*}}
\nwc{\beqn}{\begin{eqnarray*}}
\nwc{\beqast}{\begin{eqnarray*}}

\nwc{\eal}{\end{align}}
\nwc{\ee}{\end{equation}}
\nwc{\een}{\end{equation*}}
\nwc{\eea}{\end{eqnarray}}
\nwc{\eeq}{\end{eqnarray}}
\nwc{\eean}{\end{eqnarray*}}
\nwc{\eeqn}{\end{eqnarray*}}
\nwc{\eeqast}{\end{eqnarray*}}

\nwc{\invf}{\cF^{-1}_2}
\nwc{\ep}{\varepsilon}
\nwc{\tep}{\tilde{\varepsilon}}
\nwc{\epsq}{{\varepsilon^2}}
\nwc{\epsqa}{{\varepsilon^{2\alpha}}}
\nwc{\eps}{\varepsilon}
\nwc{\ept}{\epsilon}
\nwc{\vrho}{\varrho}
\nwc{\orho}{\bar\varrho}
\nwc{\ou}{\bar u}
\nwc{\vpsi}{\varpsi}
\nwc{\lamb}{\lambda}

\nwc{\nn}{\nonumber}
\nwc{\bm}{\boldmath}
\nwc{\mf}{\mathbf}
\nwc{\mb}{\mathbf}
\nwc{\ml}{\mathcal}

\nwc{\IA}{\mathbb{A}} 
\nwc{\IB}{\mathbb{B}}
\nwc{\IC}{\mathbb{C}} 
\nwc{\ID}{\mathbb{D}} 
\nwc{\IM}{\mathbb{M}} 
\nwc{\IP}{\mathbb{P}} 
\nwc{\II}{\mathbb{I}} 
\nwc{\IE}{\mathbb{E}} 
\nwc{\IF}{\mathbb{F}} 
\nwc{\IG}{\mathbb{G}} 
\nwc{\IN}{\mathbb{N}} 
\nwc{\IQ}{\mathbb{Q}} 
\nwc{\IR}{\mathbb{R}} 
\nwc{\IT}{\mathbb{T}} 
\nwc{\IZ}{\mathbb{Z}} 

\nwc{\epal}{\ep^{-2\alpha}}
\nwc{\cE}{{\ml E}}
\nwc{\cP}{{\ml P}}
\nwc{\cQ}{{\ml Q}}
\nwc{\cL}{{\ml L}}
\nwc{\cR}{{\ml R}}
\nwc{\cV}{{\ml V}}
\nwc{\cT}{{\ml T}}
\nwc{\crV}{{\ml L}_{(\delta,\rho)}}
\nwc{\cC}{{\ml C}}
\nwc{\cA}{{\ml A}}
\nwc{\cK}{{\ml K}}
\nwc{\cB}{{\ml B}}
\nwc{\cD}{{\ml D}}
\nwc{\cF}{{\ml F}}
\nwc{\cS}{{\ml S}}
\nwc{\cM}{{\ml M}}
\nwc{\cG}{{\ml G}}
\nwc{\cH}{{\ml H}}
\nwc{\bk}{{\mb k}}
\nwc{\cbz}{\overline{\cB}_z}

\nwc{\pft}{\cF^{-1}_\bp}

\begin{document}

\title{On Time Reversal Mirrors
}

\author{Albert C. Fannjiang}
 
\address{Department of Mathematics,
University of California,
Davis, CA 95616-8633}

\email{fannjiang@math.ucdavis.edu} 

\thanks{
The research is supported in part by  Darpa Grant 
 N00014-02-1-0603
}

\begin{abstract}
The concept of time reversal (TR) of scalar wave is reexamined 
 from basic principles. 

Five different time reversal mirrors (TRM)
are introduced and their relations are analyzed.
For the boundary
 behavior,  it is shown that
 for paraxial wave only the monopole TR scheme satisfies  the
exact boundary condition while for the spherical wave
only  the MD-mode TR scheme satisfies the
exact boundary condition. 

The asymptotic analysis of the  near-field focusing property
 is presented for two dimensions and three dimensions. 
It is shown that to have
a subwavelength focal spot the TRM should consist of 
 dipole transducers. The transverse resolution of
 the dipole TRM is linearly proportional to  the distance between the point source  
and  TRM.  The mixed mode TRM has  the similar (linear)
behavior  in three dimensions but in two dimensions the
transverse resolution behaves as the square-root
of the distance between the point source and TRM.  The monopole TRM is 
ineffective to focus  below wavelength.

Contrary to  the matched field processing
and the phase processor, both of  which resemble  TR, 
TR  in a weak- or non-scattering medium
is usually biased in the longitudinal direction,
especially when TR is carried out on a {\em single} plane
with a { finite} aperture.
This is true for all five TR schemes. 
On the other hand, the TR focal spot has been shown 
repeatedly  in the literature,
both theoretically and experimentally, to be centered
at the source point when the medium is multiply scattering. 
A reconciliation of the two seemingly conflicting
 results 
is found in the random fluctuations
in the intensity of the Green
function for a multiply scattering medium and the
notion of scattering-enlarged effective aperture.

\end{abstract}
\maketitle
\section{Introduction}
 
Time reversal (TR) is the process of recording the signal, time-reversing and re-propagating the signal. 
When the signal is from a localized source the time reversed
field is expected to focus on the source. 
 Time reversal of acoustic waves
 has led to
 applications in ultrasound and underwater acoustics
including
 brain therapy, lithotripsy, nondestructive testing and
 telecommunications \cite{Fin}. 
An even greater  potential holds for the time reversal
of 
 electromagnetic waves
 which is closely 
 related to optical phase conjugation  \cite{ZPS}. 
 
 Recently,  time reversal experiments with
optical and micro  waves demonstrate phase conjugation of optical near field and the robustness of 
turbidity suppression by optical phase conjugation 
\cite{Boz, BKS, RF,  LRF, YPY}. 

Motivated by these exciting experiments,
we reconsider time reversal from  basic principle,
discuss various TR schemes
and their mutual relations, 
and make general observations  about the focusing
properties of near-field and far-field  time reversal.

First, we review the principle of perfect time reversal
in a closed cavity
based on the Green second identity
and discuss the focusing properties of the perfect TR
kernel (the Porter-Bojarski kernel). We review the well known fact that
in the free space  the 
TR focal spot of a field with source
is always diffraction-limited even when all the evanescent components
are recorded  by time reversal mirror (Section \ref{sec1} and \ref{sec2}). 
The perfect time reversal is ``perfect'' only for the source-free  fields.  
On the contrary, any medium inhomogeneities, no matter
how weak, always give rise to a nonvanishing 
evanescent component in the time reversed field.

Next we derive four other  time reversal schemes (monopole, dipole and two mixed modes)  from the
Porter-Bojarski kernel   and analyze their behaviors
 with {\em one} planar TRM.  We derive their mutual  relationships
 for the paraxial wave. For the boundary
 behavior,  we show that
 for paraxial wave only the monopole TR scheme satisfies  the
exact boundary condition while for the spherical wave
only  the MD-mode TR scheme  satisfies the
exact boundary condition. 
 
A main interest here is
the near-field focusing property of various TRMs. We show that
to achieve a subwavelength focusing the near-field time reversal
mirror should involve  dipole fields  in the sense that
the TRM records not just the field but the normal  gradient
of the field and/or back-propagate  the dipole field
in proportion to the phase-conjugate recorded  data. 
The dipole TRM records and re-transmits the dipole field
while the mixed mode
TRM involves only one of the processes. The monopole TRM records
and transmits only the monopole field. 
 The transverse resolution of
 the dipole TRM is linearly proportional to  the distance between the point source  
and  TRM.  The mixed mode TRM has  the similar (linear)
behavior  in three dimensions but in two dimensions the
transverse resolution behaves as the square-root
of the distance between the point source and TRM.  The monopole TRM is 
ineffective to focus  below wavelength (Section \ref{sec3}). 

We also  point out that  TR focal spot with a finite aperture is 
generally closer to the TRM  than
the source point in the free space or
a weakly scattering medium
such as consisting of phase scatterers (Section \ref{sec4}). 
In other words, in the absence of multiple scattering
in the medium, TR as focusing or imaging method
is biased. 
In comparison, the conventional matched field
processor  and the phase processor always produce
a centered focal spot (Section \ref{sec4}).
Lastly in Section \ref{sec6}, we explain how the random effect in  multiple scattering 
can restore the centeredness of TR focal spot on the
source point as well as the
stability condition for the  robustness of 
turbidity suppression by  phase conjugation 
\cite{YPY}.

\section{Perfect time reversal}
\label{sec1}
Consider a real-valed signal in the temporal Fourier representation 
\[
U(\br,t)=\int u(\br,\omega) e^{-i\omega t} d\omega.
\]
The real-valuedness of $U$ implies that $u(\br,-\omega)=
u^*(\br,\omega)$. \commentout{
Time reversal  means reversing the direction of time in $U$, i.e.
\beqn
U(\br,-t)&=&\int u(\br,\omega) e^{i\omega t} d\omega
=\int u(\br,-\omega) e^{-i\omega t} d\omega
=\int u^*(\br,\omega) e^{-i\omega t} d\omega.
\eeqn
}
Therefore, time reversal is equivalent to phase conjugation 
of every  frequency component 
$u(\br, \omega)\to u^*(\br, \omega), \quad\forall\omega, \br. 
$
Ideally, TR turns a divergent wave into a convergent wave,
thus focusing wave energy. 
\commentout{
For instance, 
the retarded Green function for the free space
   \beq
  \label{near-free}
 G_0(\br,\br') &=&-{e^{ik|\br-\br'|}\over 4\pi |\br-\br'|}
 \eeq
becomes, after phase conjugation in the whole domain,  the advanced Green function
 \beqn
  G_0^*(\br,\br') &=&-{e^{-ik|\br-\br'|}\over 4\pi |\br-\br'|}
 \eeqn
 which is often considered as unrealistic. 
 }
 
 A practical way of
 phase conjugation  within some control domain is 
to operate TR from the boundary of  the control domain
as we consider next.

\commentout{
\begin{figure}
\begin{center}
\includegraphics[width=15cm, totalheight=6cm]{domain2.pdf}
\end{center}
\caption{Two closed cavitys: (i) a bounded domain (left),
(ii) an infinite slab (middle). The dark
areas represent medium inhomogeneities (scatterers). }
\label{domain}
 \end{figure}

\begin{figure}
\begin{center}
\includegraphics[width=15cm, totalheight=6cm]{domain22.pdf}
\end{center}
\caption{ The dark
areas represent medium inhomogeneities (scatterers). }
\label{domain}
 \end{figure}
}
\begin{figure}
\begin{center}
\includegraphics[width=15cm, totalheight=6cm]{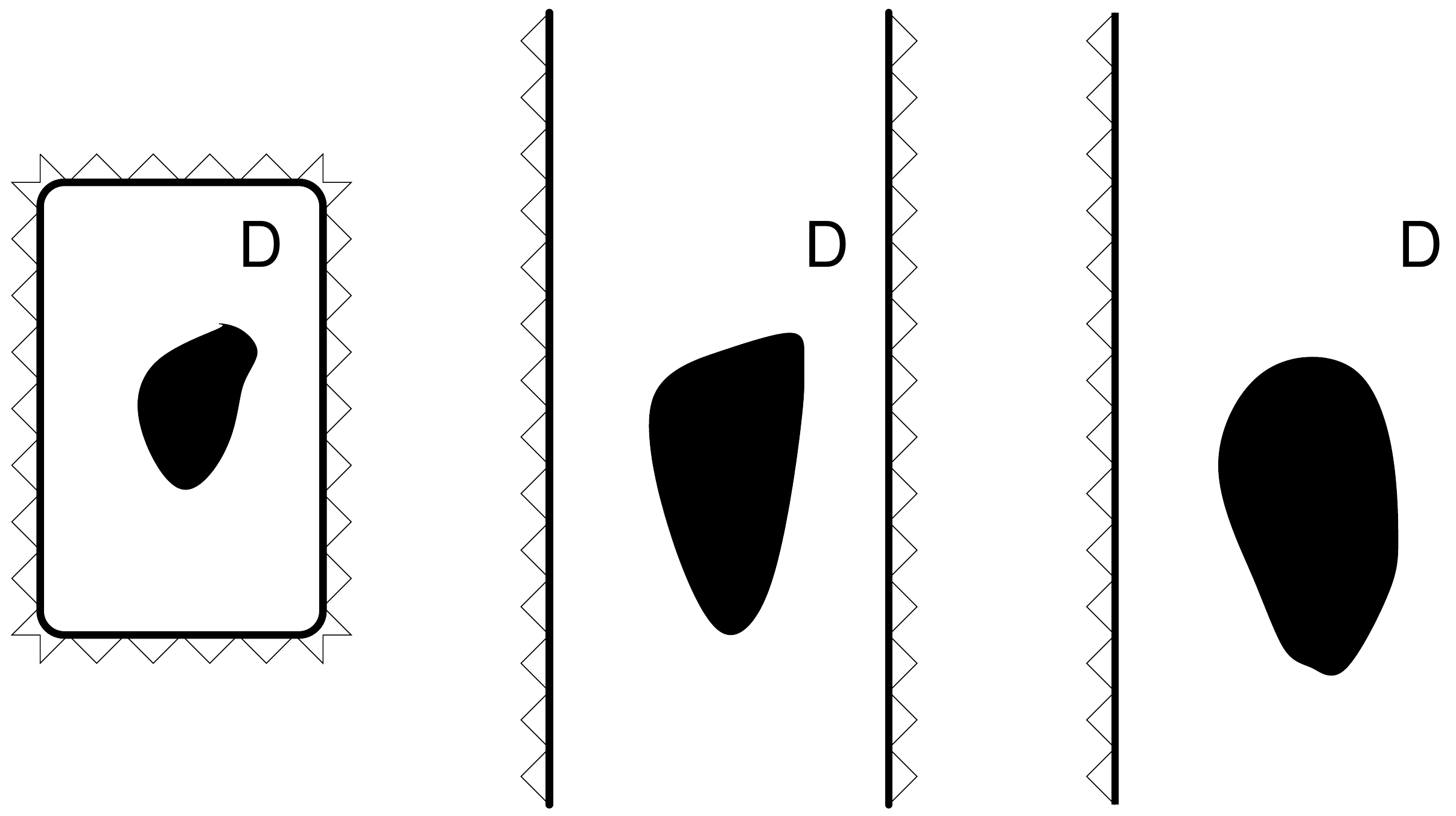}
\end{center}
\caption{Three representative domains: (i) a bounded domain (left),
(ii) an infinite slab (middle) and (iii) a half space. The dark
areas represent medium inhomogeneities (scatterers). }
\label{domain}
 \end{figure}

Consider a monochromatic scalar  wave $u$
propagating in a medium characterized by
the refractive index $n$ which in general
 varies  in space. Such a $u$ satisfies the  Helmholtz
  equation 
  \beq
  \label{helm}
  \Delta u+k^2n^2u=f
  \eeq
 subject to suitable boundary conditions
  where $f$ is the source of compact, localized  support.  
  We assume that the medium is dissipationless, i.e.
  $n^2$ is real-valued. 
 
 In 
  time reversal,   the source $f$ emits
  a wave  field $u$ which
  is then recorded at the boundary $\partial D$
  of the (topologically open) domain $D$. The 
  typical domain 
  can be one  of the following three kinds: bounded with
  a closed boundary, 
  infinite slab with two planar boundaries, or
  half space with one planar boundary (Figure \ref{domain}).

 Consider a bounded domain $D$ first.  
 Let  both $u$ and $\partial u/\partial n$ be recorded
 at $\partial D$, phase-conjugated and  back-propagated into the domain $D$ and let $v(\br)$ be the resulting wave field. Let $G(\bx,\by)$ be the Green function 
 of (\ref{helm}).  
The  time reversal principle originally proposed in 
 \cite{CF} is described by 
 \beq
 \label{green-thm}
 v(\br)\equiv \int_{\partial D} u^*(\br'){\partial G(\br, \br')\over\partial n} d\sigma(\br')
  -\int_{\partial D} {\partial u^*(\br')\over \partial n}
  G(\br, \br')d\sigma(\br'). 
 \eeq
 Clearly, $v(\br)$ is a {\em source-free} solution of the  Helmholtz equation in $D$. 

Whenever Sommerfeld's  radiation condition 
  \beq
  \label{rad}
{\partial \over \partial r}  G(\br,\br')\approx ik G(\br,\br'),
  \eeq
  for all $\br'$ sufficiently far away from $\br$, (\ref{green-thm})
  can be easily extended to the second type
  of domains (infinite slabs) through a limiting procedure. 
  Under this condition  there is no difference between 
   type (i)  (bounded)  and type (ii) (infinite
  slab) domains.

 Let $A$ be the time reversal mirror (TRM) considered
 as a part of $\partial D$. 
At each point on the boundary $\partial D \backslash A$ we impose either the sound soft boundary condition ($G=0$)
or  the
 sound hard boundary condition ($\partial G/\partial n=0$).  
The boundary condition on TRM $A$ is flexible 
 except  that the presence of TRM does not
affect the near-field behavior  of the Green function, see
(\ref{near2})-(\ref{near3}).
 We call either  type (i) or (ii) domain of Figure \ref{domain}
 {\em closed cavity}.
 \commentout{ if either the sound soft
 or the sound hard condition is satisfied on $\partial D\backslash A$. Type (iii) domain (half space) in Figure \ref{domain} 
 can be obtained through the limiting procedure
 from  type (ii) domains if the perfectly matched 
 condition is imposed on the right boundary. 
 }

 \commentout{
The second Green theorem
   \beq
 && \int_{\partial D} u^*(\br'){\partial G(\br, \br')\over\partial n} d\sigma(\br')
  -\int_{\partial D} {\partial u^*(\br')\over \partial n}
  G(\br, \br')d\sigma(\br')\nn\\
  & =&\int_D u^*(\br')\Delta G
  (\br, \br') d\br'
  -\int_D \Delta u^*(\br') G(\br, \br')d\br'\label{green2}
  \eeq
The right hand side of (\ref{green2})
 should be interpreted in the sense of distribution for $\br \in D$.
 
 Adding and subtracting the term $u^*(\br')k^2n^2(\br')G
  (\br, \br')$ to the right hand side of (\ref{green2}) we have 
  }
 By  the second Green's identity  we have
   \beq
 && \int_{\partial D} u^*(\br'){\partial G(\br, \br')\over\partial n} d\sigma(\br')
  -\int_{\partial D} {\partial u^*(\br')\over \partial n}
  G(\br, \br')d\sigma(\br')\nn\\
  & =&\int_D u^*(\br')(\Delta+k^2n^2(\br'))G
  (\br, \br') d\br'
  -\int_D (\Delta +k^2n^2(\br')) u^*(\br') G(\br,\br')d\br'.\label{green3}
  \eeq
The time reversal operation described
by  (\ref{green-thm}) is  perfect
for 
 the field {\em inside} $D$ when the field
 is  source-free inside 
$D$ (i.e.  $u$ satisfies 
(\ref{helm}) with $f|_D=0$ and the second integral on the right hand side of (\ref{green3})  drops out)
  \beq
  \label{out2}
 v(\br)=
 u^*(\br),\quad \br\in D
 \eeq
 while $v(\br)=0, \forall \br \notin \bar D$ by (\ref{green3}).
 Hence (\ref{green-thm}) has been proposed as  the basis
for time reversal in a closed cavity \cite{CF, RF}. 

\commentout{
On the other hand, it is known that
phase conjugation on one plane as in type (iii) domain
(Figure \ref{domain}) may not produce point-wise phase conjugation in the whole domain
(see \cite{NW}). 
}

On the other hand, 
if $D$ contains sources  as in  typical applications,  then the time reversal described by (\ref{green-thm}) is diffraction-limited. Indeed, 
\commentout{
By (\ref{helm}) and (\ref{green3}), we have 
  \beq
  \label{green4}
 v(\br)&=& u^*(\br)-\int_D f^*(\br')G(\br, \br')d\br', \quad
 \br\in D
 \eeq
 and 
 \beq
 \label{outside}
 v(\br)=-\int_D f^*(\br')G(\br, \br')d\br', \quad
 \br\notin \bar D. 
 \eeq
We also have by the definition of the Green function that
\beq
\label{uc}
 u(\br)&=&\int_D G(\br, \br')f(\br')d\br'
 \eeq
and hence
}
  \beq
 \label{green4}\label{tr-sinc}
 v(\br)&=& u^*(\br)-\int_D f^*(\br')G(\br, \br')d\br'\\
  &=&\int_D \lt[G^*(\br, \br')-G(\br, \br')\rt]
  f^*(\br')d\br'\nn\\
 &=&\int_D K_{\rm PB}(\br,\br')f^*(\br')d\br',\quad\br\in D\nn
  \eeq
with the Porter-Bojarski kernel  
  \beq
  \label{PB}
  K_{\rm PB}(\br,\br')=-2\Im[G(\br,\br')] 
  \eeq
  where $\Im$ denotes
  the imaginary part, c.f. \cite{CF}.
  With a real-valued $n^2$, $K_{\rm PB}(\br,\br')$ is
  an example of a source-free radiation field in $\br$ for each $\br'$. In the free space, (\ref{PB}) is always diffraction-limited
  (see Section \ref{sec2}). 
  

The reason that $K_{\rm PB}\neq G^*$ is because
the time symmetry is broken when the field
is phase conjugated but not the source. 
A time-reversed source is a sink. 
  Suppose now the point source  becomes  a  point sink 
  after emitting the initial  wave and suppose the sink
  can absorb the time-reversed, incoming wave and 
  prevent the re-emission  of the outgoing wave
 with efficiency $\rho>0$, \cite{RF2}. 
  Then (\ref{green4}) becomes
    \beq
  \label{green-sink}
 v(\br)
 &=&G^*(\br,\br')-(1-\rho)G(\br,\br_0),\quad\br\in D,
 \eeq
which, denoted by $K_{\rm S}$,  is   the TR kernel with a point sink of efficiency $\rho$. When $\rho=1$ then $K_{\rm S}=
G^*$. As we shall see in Section \ref{sec2},
 the TR focusing property  for $K_{\rm S}$  
 is  infinitely enhanced over  that for (\ref{PB}).

\commentout{Needless to say,  in practice,  a perfect point sink
is technically difficult to realize (see \cite{RF2} for
acoustic sink and \cite{Kot1, Kot2} for
  theory and experiment of black screen which is 
  the optical analogy 
  to the acoustic  sink).
  }
  
\commentout{
 \beq
 \label{tr-sink}
 K_{\rm S}(\br,\br')=(1-\rho)G^*(\br, \br')-G(\br, \br').
 \eeq
 }
 
 \section{Monopole and dipole  TRMs}
 In this section we derive approximate TR schemes  
 from (\ref{green-thm}) using Sommerfeld's radiation
 condition (\ref{rad}).
 
  \label{sec3}
For a closed cavity, using 
the sound soft/hard condition on  $\partial D\backslash A$, we write
     \beq
\lefteqn{ \int_{\partial D} u^*(\br'){\partial G(\br, \br')\over\partial n} d\sigma(\br')
  -\int_{\partial D} {\partial u^*(\br')\over \partial n}
  G(\br, \br')d\sigma(\br')}\nn\\
  &=&\int_D \lt[\int_{A} G^*(\br', \br''){\partial G(\br, \br')\over\partial n} d\sigma(\br')
  -\int_{A} {\partial G^*(\br', \br'')\over \partial n}
  G(\br, \br')d\sigma(\br')\rt] f^*(\br'')d\br''\nn
  \eeq
  which yields the alternative expression for 
  the Porter-Bojarski kernel for a closed cavity
  \beq
  \label{PB2}
  K_{\rm PB}(\br,\br'')&=&\int_{A} G^*(\br', \br''){\partial G(\br, \br')\over\partial n} d\sigma(\br')
  -\int_{A} {\partial G^*(\br', \br'')\over \partial n}
  G(\br, \br')d\sigma(\br')
  \eeq

 Now  we pretend $|\br'-\br''|, |\br-\br'|\gg 1$ 
and   use condition (\ref{rad}) to motivate two other TR schemes. 
  We consider 
all three types of domains depicted in Figure
 \ref{domain}. 

Replacing  $
  \partial G/\partial r$ with $ikG$ or   $ikG$ with
  $\partial G/\partial r$ in (\ref{PB2}) leads to, respectively, 
the monopole TR kernel
  \beq
  \label{LB}
  K_{\rm M}(\br,\br'')&=&-
  2ik\int_{A} G^*(\br',\br'')G(\br,\br')d\sigma(\br')
  \eeq
  and 
  the dipole TR kernel
    \beq
  \label{DP}
  K_{\rm D}(\br,\br'')&=&-{2i\over k}\int_{A} {\partial \over\partial n} G^*(\br', \br''){\partial \over\partial n}G(\br, \br') d\sigma(\br'). 
  \eeq

In addition, we introduce mixed mode TR kernels
\beq
\label{MD}
K_{\rm MD}(\br,\br'')&=&\int_{A} G^*(\br', \br''){\partial G(\br, \br')\over\partial n} d\sigma(\br')\\
K_{\rm DM}(\br,\br'')&=&-\int_A {\partial G^*(\br', \br'')\over \partial n}
  G(\br, \br')d\sigma(\br')\label{DM}
  \eeq
 which are parts of (\ref{PB2}).

 Analogously we define the various TR schemes for
a source-free field $u$:
\beq
  \label{LB'}
  v_{\rm M}(\br)&=&-
  2ik\int_{A} u^*(\br')G(\br,\br')d\sigma(\br')\\
    \label{DP'}
  v_{\rm D}(\br)&=&-{2i\over k}\int_{A} {\partial \over\partial n} u^*(\br'){\partial \over\partial n}G(\br, \br') d\sigma(\br')\\
\label{MD'}
v_{\rm MD}(\br)&=&\int_{A} u^*(\br'){\partial G(\br, \br')\over\partial n} d\sigma(\br')\\
v_{\rm DM}(\br)&=&-\int_A {\partial u^*(\br')\over \partial n}
  G(\br, \br')d\sigma(\br')\label{DM'}\\
  v_{\rm PB}(\br)&=&v_{\rm MD}(\br)+v_{\rm DM}(\br)\label{PB'}
  \eeq
  
{  The various TR schemes can be readily implemented  for the acoustic wave. For the dipole TRM, the transducers record  the
normal pressure gradient and emit the normal dipole field in proportion
to the phase-conjugate recorded data. 
For the mixed mode TRM, the transducers
record the pressure (resp. pressure gradient) and emit
the normal dipole (resp. monopole)  field in proportion to the phase-conjugate
recorded data (see, for example, \cite{RTLF} for a similar proposal).  
}

 The time reversal schemes
 described by  (\ref{PB2})-(\ref{DM})
 and (\ref{LB'})-(\ref{PB'})  are
 main objects of subsequent analysis. 
 Note that the expressions (\ref{PB2})-(\ref{DM}),
 (\ref{LB'})-(\ref{PB'}) are source-free fields in the domain
 $D$ regardless whether the initial wave is source-free or not.

\subsection{Paraxial wave}
In the case of paraxial wave, (\ref{PB2}),  (\ref{LB}), (\ref{MD}) and
(\ref{DM}) are explicitly connected with one another as follows.

 Consider 
 the half space $D=\{z>0\}$ with
a  planar TRM on the transverse plane $z=0$. Let $G(z,\bx,z',\bx')$ be
the paraxial Green function 
where $z,z'$ are the longitudinal coordinate and 
$\bx,\bx'$ are the transverse coordinates (see Appendix A). 
Let $\br_0=(z_0,\bx_0)$ be the location of
the source. 
 Assume that the TRM is away from the medium inhomogeneities
  i.e. $n(0,\bx)=1$. Then it can be shown
  that 
\beq
\label{38} K_{\rm MD}(z,\bx, z_0,\bx_0) &=&{1\over 2} K_{\rm M}(z,\bx, z_0,\bx_0) +{i\over 2k}\int \nabla'_{\rm T} G^*(z_0,\bx_0,0,\bx') \cdot\nabla'_{\rm T} G(z,\bx,0,\bx')d\bx'\\
 &=& K_{\rm DM}(z,\bx, z_0,\bx_0)
  \label{39}\nn
\eeq
where $\nabla'_{\rm T}$ denotes the transverse gradient w.r.t. $\bx'$ (Appendix A).
Therefore,
\beq
\label{mono2}
K_{\rm MD}(\br,\br_0) &=&K_{\rm DM}(\br,\br_0)=
{1\over 2} K_{\rm PB}(\br,\br_0). 
\eeq

\commentout{
 The two terms  on the right hand side of (\ref{38}) or
(\ref{39})  have different natures:  the first term typically
dominates in the far field while the second
term dominates in the near field. 
}

Likewise, we have
\beqn
\label{MD'''}\nn
v_{\rm MD}(z,\bx)&=&{1\over 2} v_{\rm M}(z,\bx)+{i\over 2k} \int\nabla_{\rm T} u^*(0,\bx')\cdot\nabla_{\rm T} G(z,\bx,0,\bx')d\bx'\\
&=&
v_{\rm DM}(z,\bx) ={1\over 2} v_{\rm PB}(z,\bx).
  \eeqn
 
\subsection{Boundary value}
Though  not fundamental to many time reversal
 applications,
it is sometimes desirable  for the TR schemes to  satisfy  the  boundary condition
 \beq
 \label{BC}
K_{\rm X} (\br,\br_0)=G^*(\br,\br_0),\quad
 v_{\rm X}(\br)=u^*(\br), 
 \quad \br\in A
 \eeq
 where $\quad {\rm X}={\rm M, D, MD, DM, PB}$. 
We have already seen that in a closed cavity $v_{\rm PB}$ satisfies the boundary condition  but $K_{\rm PB}$ does not.

In Appendix \ref{app:bc} we show that, surprisingly, 
{\em  for the paraxial wave, only the monopole TR scheme satisfies the boundary condition (\ref{BC}) and
for the spherical wave in the {free} space, 
only the mixed mode TR scheme $v_{\rm MD},  K_{\rm MD}$,  multiplied by $2$,  satisfy  (\ref{BC})}. 
\commentout{However, this nice property with the MD-mode TR is an exception rather
than a rule. It fails to hold for example for a scattering medium where
the refractive index depends on
the longitudinal coordinate or the transverse coordinates
}


  \section{Near-Field focusing}
  \label{sec2}
  In this section, we analyze  the near-field focusing property of various TR schemes.
We show that
 $K_{\rm D}, K_{\rm MD}, K_{\rm DM}$
give rise to a subwavelength focal spot of size
proportional to the distance from  the source to 
TRM while $K_{\rm M}$ and $K_{\rm PB}$ can hardly
achieve subwavelength focusing. 
In the free space,
the focal spot
size for the paraxial wave is always comparable
to the Fresnel length $\sqrt{\lambda z_0}$ which
is not applicable  for $z_0\ll \lambda$.

Consider the Porter-Bojarski kernel for
 the free space  Green function in three dimensions, 
 \beq
  \label{near}
K_{\rm PB} &=&{\sin{(k|\br-\br'|)}\over 2\pi |\br-\br'|}
  \eeq
 and compare it with the real part of  $K_{\rm S}$ defined by
  (\ref{green-sink}) 
  \beq
    \label{near22}
\Re[K_{\rm S}]&= &\rho{\cos{(k|\br-\br'|)}\over 4\pi |\br-\br'|},\quad\rho>0
 \eeq
 which dominates over the imaginary part 
for $\rho\approx 1$  and $|\br-\br'|\leq\lambda/4$.
 \commentout{
 as the point-spread function of a hypothetical imaging system
  from the perspective of resolution. Indeed, in Wheeler-Feynman's  time
  symmetric formulation of classical electrodynamics 
  \cite{WF1, WF2},
  (\ref{near22}) is the Green function of a point emitter in the free space.

  Note that (\ref{near22}) is $-1/\rho$ times 
 the real part of $K_{\rm S}$ when the
 medium is  the free space and the TRM $A=\partial D$ is
a perfect absorber. 
}

Resolution of an imaging system  generally refers to either the focal spot size
or the minimum resolvable distance between two points or lines (two-point or -line resolution). And these two ideas are related: the focal
spot size is a crude estimate of  the minimum  resolvable
distance.  
 
 With all the insights into  superresolution techniques, 
  it is now well accepted that 
  in the absence of noise and with perfect knowledge of
  the imaging system there is no limitation to
  the two-point resolution.
 Fundamentally,
  this is because an image with infinite signal-to-noise
  ratio can convey  an arbitrary  amount of information
  \cite{BB}. 
  
  Since we do not account for the noise explicitly 
  we will focus on the idea of focal spot size as a measure
  of focusing and image sharpness. 
   According to Rayleigh's criterion which essentially
  defines the spot size as the distance of the first zeros
  to the maximum point, (\ref{near}) has
  a spot size of $\lambda/2$ and (\ref{near22}) has
  a spot size of $\lambda/4$. But in reality 
  the focal spot is  in some sense ``in the eye of the viewer''
  and one readily recognizes
   the vastly sharper  graph  of
  (\ref{near22}) than that of (\ref{near}).  
  
  This contrast  is better described
  by the standard  (Houston) criterion used in astronomy
which  is to adopt the ``full width at half maximum'' (FWHM) \cite{Hou}. The FWHM of (\ref{near}) is about $\lambda/2$
while the FWHM of (\ref{near22}) is zero since
the maximum of (\ref{near22}) is infinity (therefore
infinitely sharp). Indeed,
the singularity is the
signature  of a point  source. 

\commentout{
\begin{figure}
\begin{center}
\includegraphics[width=7cm, totalheight=6cm]{path.pdf}
\end{center}
\caption{The paths of integration $C^+, C^-$ in the complex $\alpha$-plane. }
\label{fig:path}
 \end{figure}
}

The subwavelength, superresolution of (\ref{near22}) can be understood from Weyl's angular spectrum of plane waves.
For  the imaging  direction $z$, the free space Green function in three
dimensions  has the angular spectrum 
representation 
\beq
{e^{ikr}\over r}={ik\over 2\pi}
\int {ds_1ds_2\over s_3} 
 \exp{\lt[ik(s_1(x-x_0)+s_2(y-y_0)+
 s_3|z-z_0|)\rt]}\label{weyl}
 \eeq
 where
 \beqn
 r&=&\sqrt{(x-x_0)^2+(y-y_0)^2+(z-z_0)^2}\\
s_3&=&\sqrt{1-s_1^2-s_2^2},\quad s_1^2+s_2^2\leq 1\\
s_3&=&i\sqrt{s_1^2+s_2^2-1},\quad s_1^2+s_2^2>1
\eeqn
\cite{Cle}. 

\commentout{
\beq
\label{weyl}
{e^{ikr}\over r}&=&{ik\over 2\pi}
\int^\pi_{-\pi}d\beta \int_{C^\pm}d\alpha 
\cos{\alpha} \exp{\lt[ik\bs\cdot \br\rt]}
\eeq
where 
$\bs$ is a unit vector parametrized by the polar angles
$\alpha, \beta$ as 
\beq
\label{17}
s_1=\cos{\alpha}\cos{\beta},\quad s_2=\cos{\alpha}\sin{\beta},
\quad s_3=\sin{\alpha}.
\eeq
For $z>0$, the integration is over the path 
$C^+$  oriented clockwise from the point $(\pi/2, 0)$ to the origin
along the real axis and from the origin to 
$i\infty$ along the positive imaginary axis; for $z<0$ the integration is over
the path 
$C^-$ oriented counterclockwise 
from $-i\infty$ to the origin along the negative
imaginary axis and from the origin  to the point $(-\pi/2,0)$ along
the real axis \cite{BW, Cle}. 

The integrand in (\ref{weyl}) with  real-valued $\alpha$
corresponds to the propagating (homogeneous) wave
and that with purely imaginary $\alpha$ corresponds
to the evanescent (inhomogeneous) wave. 
}
The integrand in (\ref{weyl}) with $s_1^2+s_2^2\leq 1$
corresponds to the propagating (homogeneous) wave
and that with $s_1^2+s_2^2>1$  corresponds
to the evanescent (inhomogeneous) wave. 
Let $W_{\rm H}$ and $W_{\rm E}$ denote
the homogeneous and the evanescent fields, respectively.
Since $k\sqrt{s_1^2+s_2^2}$ is the transverse wavenumber, 
$W_{\rm H}$ contains the diffraction-limited information
about the point source while  $W_{\rm E}$ contains
the sub-wavelength information about the point source.  
\commentout{
\beq
\label{weyl}
{e^{ikr}\over r}&=&{ik\over 2\pi}
\int \exp{\lt[ik(px+qy+m|z|)\rt]}\frac{dpdq}{m}
\eeq
where 
\beq
\label{homo} m&=&\sqrt{1-p^2-q^2},\quad p^2+q^2\leq 1\\
\label{evan} m&=&i\sqrt{p^2+q^2-1},\quad p^2+q^2>1.
\eeq

The integration in (\ref{weyl}) over (\ref{homo})
corresponds to the so called homogeneous wave $W_{\rm H}$;
the integration in (\ref{weyl}) over (\ref{evan})
corresponds to the evanescent wave denoted by $W_{\rm E}$.
}

The first important observation is that
 $W_{\rm E}$ is always real-valued.
 Consequently, the TR kernel $K_{\rm PB}$ in the case of
 $G=G_0$ contains
 no  evanescent component. 
 Secondly, 
 $W_{\rm H}$ is bounded
and in the $\delta$-neighborhood of the source point
 the estimate holds:
$|\Re[W_{\rm H}]|=O(\delta)$. In other words,
in the immediate vicinity  of the source point
$\Re[G_0]$ essentially coincides with  $W_{\rm E}$. 
This means that $\Re[G_0]$ contains all the subwavelength
information about the point source.

Therefore, 
  (\ref{tr-sinc}) is always diffraction-limited
  no matter how narrow the support of $f$ is and
  how close the source is to the TRM. 
This means that in the absence of scattering
even when the evanescent
  waves are recorded, phase-conjugated and
  back-propagated, the resulting field is still smeared 
 below the scale of  wavelength. In general,  if $|\Re[G]|\gg |\Im[G]|$ in the near field as is the case for the free space,  the latter
 does not contribute to the leading order effect
 of the near field TR and  hence that TR, which affects 
 only the phase,  has
 a secondary effect on 
 the subwavelength resolution of the near field TR. 

 

 Since we are concerned  here with the asymptotic behavior
of near field TR as the point source approaches
the TRM, we  relax the definition of FWHM.  Instead of
 ``full width at half maximum'' in the literal sense  we 
 also refer to the (asymptotic)  resolution as 
the asymptotic size of a focal spot 
as the maximum width at the same 
(or the minimum width at a slower) asymptotic behavior
of the point-spread  function. For example, 
in  this  metric,  the focal spot size for (\ref{near}) 
is $\lambda/2$ while the truncated version of (\ref{near22}):
\beq
\label{cutoff}
\min\lt\{ {1\over \delta}, \rho
{\cos{(k|\br-\br'|)}\over 4\pi |\br-\br'|}\rt\},\quad\delta\ll 1
\eeq
has an asymptotic spot size $O(\delta)$.

Next we shall analyze the near-field focusing property
of the various TR schemes. 
We assume the near field asymptotic 
   \beq
  \label{near2}
  |G(\br,\br')| &\sim &{1\over  |\br-\br'|},\quad |\br-\br'|\leq\lambda.
 \eeq
For $|\br-\br'|\ll\lambda$ (\ref{near2}) is the
generally correct asymptotic 
as the near field singularity 
is  determined solely by the Laplacian  in the Helmholtz equation \cite{Joh}.  
In addition to (\ref{near2}) we further assume
  \beq
  \label{near3}
\nabla  G(\br,\br') &\sim &{\br-\br'\over |\br-\br'|^3},
\quad |\br-\br'|\leq \lambda.
\eeq

\subsection{Transverse resolution  of
monopole TR}

Consider the case of  a point monopole source located at
$\br_0$ in close proximity of  $A$. Let  $\br'_0$ be the closest point on $A$ to $\br_0$ and $|\br_0-\br_0'|=\delta\ll 1$.
Using (\ref{near2}) we have the estimate
\[
K_{\rm M}(\br_0, \br_0)=O(-k\log{\delta})
\]
 since the singularity in 
the integrand of (\ref{LB}) is 
\beq
\label{sing1}
{k\over |\br'-\br_0||\br-\br'|},\quad |\br-\br_0|\ll \lambda,
\quad |\br'-\br'_0|\ll \lambda. 
\eeq
More explicitly, the integral (\ref{LB}) can be
simplified by replacing the surface integral
over $A$ by the tangent plane at $\br_0'$. Without
loss of generality, we may assume the tangent plane to be
the $x-y$ plane and $\br_0=(0,0,\delta)$. 
The leading order transverse profile of $K_{\rm M}$ around $\br_0$
in  $x$  is given by
\beq
\label{20}
k\int { (x'^2+y'^2+\delta^2)^{-1/2}}
{( (x-x')^2+y'^2+\delta^2)^{-1/2}}dx'dy',\quad\delta\ll 1
\eeq
where the integration is restricted to a bounded area, say
the circular disk of radius $\lambda$, 
around the origin to avoid
logarithmic divergence at large $x', y'$. 

Let $x=\delta^\alpha, 0\leq \alpha<1$. Then (\ref{20}) has
the same asymptotic as
\beq
\label{20'}
k\int { (x'^2+y'^2)^{-1/2}}
{( (\delta^\alpha-x')^2+y'^2)^{-1/2}}dx'dy',\quad\delta\ll 1
\eeq
where the integration is over the punctured  disk  with holes
of radius $\delta$ centered at $(0,0)$
and $(\delta^\alpha, 0)$ since the two holes
have a vanishing contribution to (\ref{20}). The change of  variables from
$x', y'$ to $x''=x'/\delta^\alpha, y''=y'/\delta^\alpha$
renders (\ref{20'}) into
\beq
\label{20''}
k\int { (x''^2+y''^2)^{-1/2}}
{( (1-x'')^2+y''^2)^{-1/2}}dx''dy''
\eeq
integrated over  the punctured disk of
radius $\lambda\delta^{-\alpha}$ with shrinking holes
of radius $\delta^{1-\alpha}$ centered at
$(0,0)$ and $(1,0)$. 
Hence
\beq
\label{41-2}
K_{\rm M}(\br,\br_0)\sim 
\lt\{\begin{matrix}
-k\alpha\log \delta,&\alpha\in (0,1)\\
k,&\alpha=0
\end{matrix}\rt.,\quad\delta\ll 1 
\eeq
where the logarithmic divergence for $\alpha>0$ is
due to the integration (\ref{20''}) on the scale of $\delta^{-\alpha}$. \commentout{
This implies that the focal spot size is larger than
$\delta^\alpha$ for any $\alpha>0$.
  The same analysis
with $x=(-1/\log{\delta})^\beta, \beta>0, $ reveals that $K_{\rm M}(\br,\br_0)
=O\lt(\beta\log{\lt(-\log{\delta}\rt)}\rt)$ and hence 
the asymptotic focal spot size is smaller than  $(-1/\log{\delta})^\beta, \forall \beta>0$.
}

\begin{figure}
\begin{center}
\includegraphics[width=8cm, totalheight=6cm]{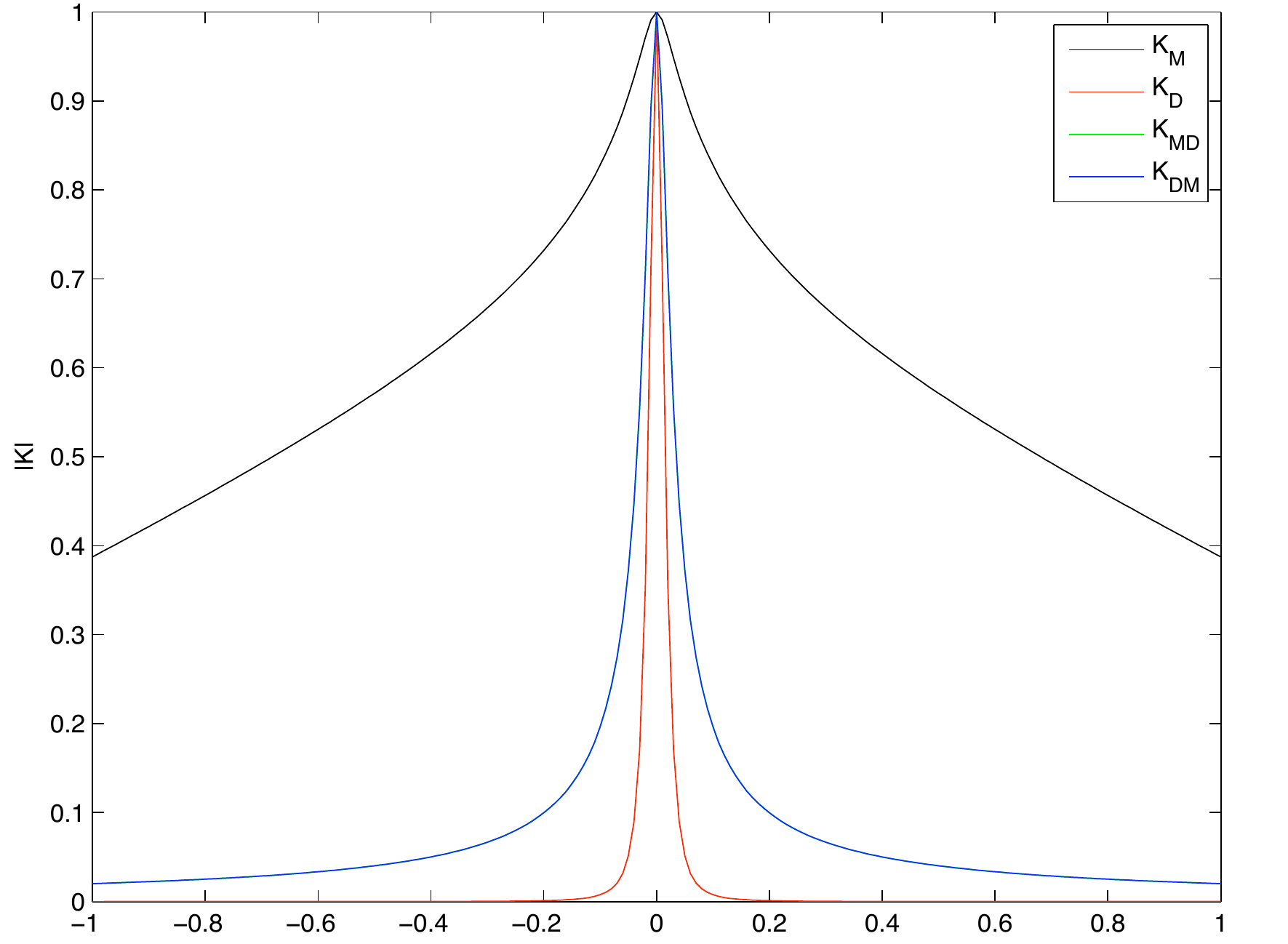}
\includegraphics[width=8cm, totalheight=6cm]{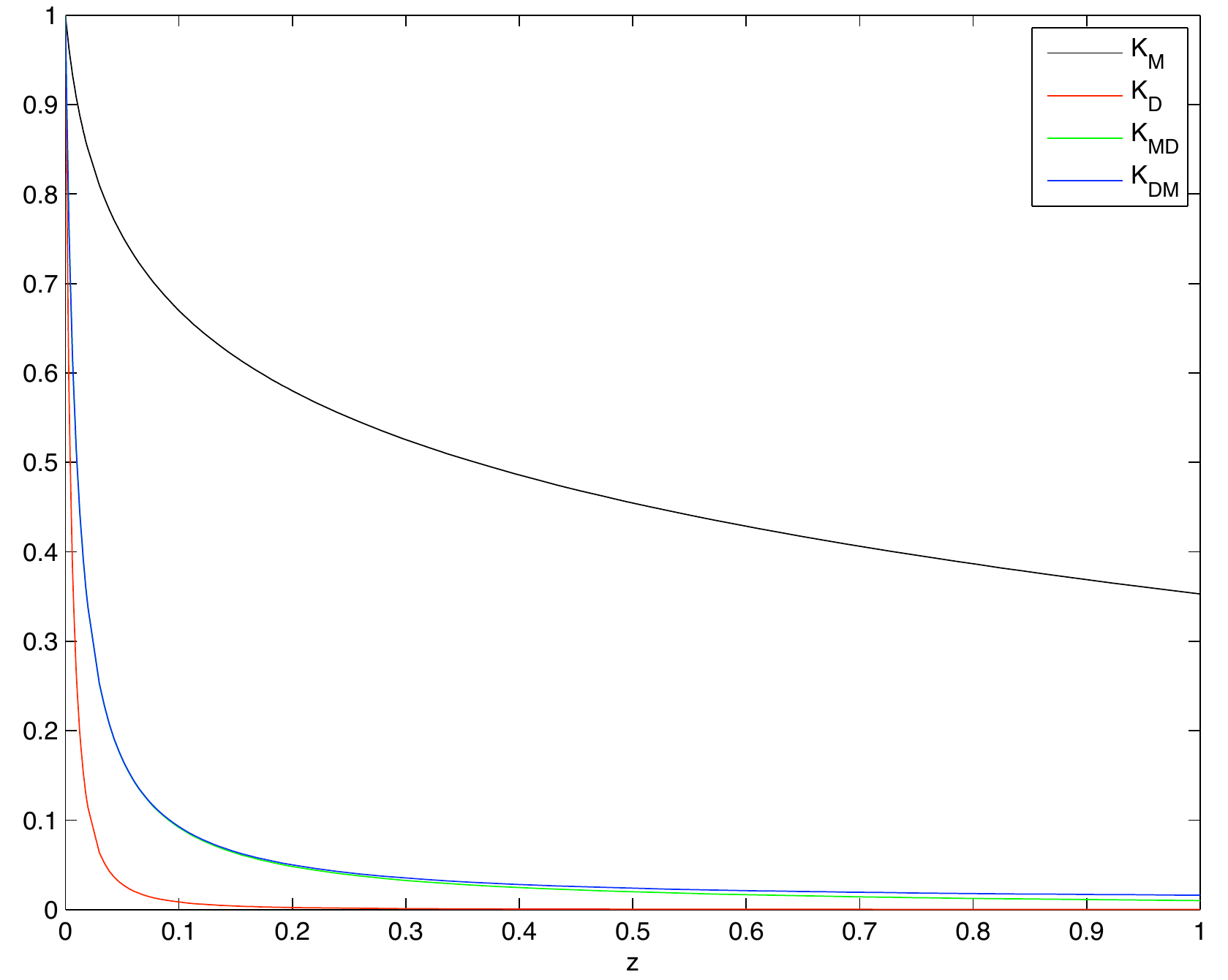}
\end{center}
\caption{(Left) 3d transverse   profiles with FWHM $=1.37 (K_{\rm M}), 0.03 (K_{\rm D}), 0.07 (K_{\rm MD}, K_{\rm DM})$;
(Right) 3d longitudinal profiles with FWHM $=0.3565(K_{\rm M}) , 0.00421 (K_{\rm D}), 0.00947 (K_{\rm DM}, K_{\rm MD}).$  Note that the transverse profiles of 
the $K_{\rm MD}$ and $K_{\rm DM}$ coincide. }
\label{fig:3d}
 \end{figure}

When $\delta$ is so  small  that $\log{\delta}$ is
the dominant behavior of  $K_{\rm M}$ near the source point, (\ref{41-2}) says
that at $x=\delta^{1/2}$,  $K_{\rm M}$ is roughly
at half the maximum and therefore the transverse
resolution  is proportional to $\delta^{1/2}$. 
However, this scaling behavior requires extremely small $\delta$
to manifest itself. Outside this regime, the focal spot size would
decrease slowly as $\delta$ decreases (see 
the comments on Figure \ref{fig:3d}). 
\commentout{
Calculating the derivative of $K_{\rm M}$ is more subtle 
and can be simplified by replacing the surface integral
over $A$ by the tangent plane at $\br_0'$. Without
loss of generality, we may assume the tangent plane to be
the $x-y$ plane and $\br_0=(0,0,\delta)$. 
Differentiating $K_{\rm M}$ with respect to $x$ we have the expression
\[
\int {1\over (x'^2+y'^2+\delta^2)^{1/2}}\times
{x-x'\over ( (x-x')^2+y'^2+\delta^2)^{3/2}}dx'dy'
\]
which vanishes at $x=0$ due to the anti-symmetry
of the integrand with respect to $x'$ indicating
that $\br_0$ is a local maximum in $x$.

Therefore, the monopole TR can hardly 
achieve arbitrarily fine resolution even as 
 the point source approaches  the TRM.
}
 
 \subsection{Transverse resolution of dipole TR }

For  $\br'= \br_0'$ and $\br=\br_0$, $\br-\br'$ is 
orthogonal to $A$ and thus  $|\nabla G|\sim |\partial G/\partial n|$. Therefore the leading order transverse profile
 for $K_{\rm D}(\br,\br_0)$ is given approximately by 
\beq
\label{24}
{1\over k}\int {\delta\over (x'^2+y'^2+\delta^2)^{3/2}}\times
{\delta\over ((x-x')^2+y'^2+\delta^2)^{3/2}}dx'dy'.
\eeq
The integrand  in (\ref{24}) has
a different singularity than that in (\ref{20}). Indeed, for $x=\delta^\alpha$,  the
dominant contribution to (\ref{24}) comes from
the disks of radius $\delta$ around $(0,0)$ and
$(\delta^\alpha, 0)$ which is easily estimated to be
\[
K_{\rm D}(\br,\br_0)\sim \delta^{1-3\alpha},
\quad\alpha\in [0,1].  
\]
By contrast, $K_{\rm D}(\br_0,\br_0)\sim \delta^{-2}$. 
Hence  the focal spot size in this case is smaller
than 
$\delta^\alpha, \forall \alpha\in [0, 1)$ asymptotically.
We conclude that the focal spot size for the dipole TR is
essentially $O(\delta)$
 and that  the dipole TR can in principle break the
diffraction limit to achieve an arbitrarily fine
resolution by bring the TRM sufficiently close to the source.

\commentout{
As a result, the singularity of the integrand in  (\ref{DP}) 
is 
\beq
\label{sing2}
{1\over k|\br'-\br_0|^2|\br-\br'|^2},\quad |\br-\br_0|\ll \lambda,
\quad |\br'-\br'_0|\ll \lambda
\eeq
which after integrating over  $A$ scales like $\lambda\delta^{-2}$.

 Differentiating $K_{\rm D}$ with respect to $x$ we have
\[
\int {\delta^2\over (x'^2+y'^2+\delta^2)^{3/2}}\times
{x-x'\over ((x-x')^2+y'^2+\delta^2)^{5/2}}dx'dy'
\]
}

\commentout{
This superresolution effect has been observed
in experiments with phase conjugation of  optical near field \cite{VB}.
The actual physical process involved is more complicated
than what the expression (\ref{DP}) describes. 
}

  \commentout{
  
  In particular, for a point source $f(\br)=\delta(\br-\br_0)$, we
  have
  \[
  v(\br)={1\over 2\pi i} {\sin{(k|\br-\br_0|)}\over |\br-\br_0|}.
  \]
  In this case, the time reversal resolution is proportional to $\lambda$.    Note that (\ref{tr-sinc}) is independent of the domain $D$
  as long as it contains the support of $f$. 

  }
   
\subsection{Transverse resolution of mixed mode TR}
\commentout{
Similar argument as above shows
that  the singularity of the integrands in
(\ref{MD}) and (\ref{DM})  are, respectively,  
\beqn
{1\over |\br'-\br_0||\br-\br'|^2},&& \quad |\br-\br_0|\ll \lambda,
\quad |\br'-\br'_0|\ll \lambda \\
{-1\over |\br'-\br_0|^2|\br-\br'|},&&\quad |\br-\br_0|\ll \lambda,
\quad |\br'-\br'_0|\ll \lambda 
\eeqn
which after integrating over  $A$ scales like $\delta^{-1}$. 
}
By the same analysis as before, the integrals
(\ref{MD}) and (\ref{DM}) can be approximated by, respectively,
\beq
\label{26}
&&\int {1\over (x'^2+y'^2+\delta^2)^{1/2}}\times
{\delta\over ((x-x')^2+y'^2+\delta^2)^{3/2}}dx'dy'\\
&&-\int {\delta\over (x'^2+y'^2+\delta^2)^{3/2}}\times
{1\over ((x-x')^2+y'^2+\delta^2)^{1/2}}dx'dy'\label{27}. 
\eeq
Set $x=\delta^\alpha, 0\leq \alpha<1$. In the case of (\ref{26}), the dominant contribution
($\sim \delta^{-\alpha}$)
comes from integration around $(\delta^\alpha, 0)$ for
$\alpha<1$ while, in the case of (\ref{27}), the
dominant contribution ($\sim \delta^{-\alpha}$) 
comes from integration around $(0,0)$. 
In both cases $K_{\rm MD}(\br_0,\br_0)
\sim K_{\rm DM}(\br_0,\br_0)\sim \delta^{-1}$. 
Hence the focal spot size is smaller than
$\delta^\alpha,\forall \alpha\in [0,1)$ asymptotically. 

We conclude that the focal spot size for the mixed mode TR
is essentially $O(\delta)$ and that the mixed mode TR
can achieve an arbitrarily fine resolution as the TRM approaches
the 
point source. Moreover, the fact that
the maximum values of the point-spread functions
are inversely proportional to 
 the focal spot size is consistent with
 the cut-off version of (\ref{near22}) in the following sense.
 Truncate the Green function (\ref{near2}) at the level
 $\delta^{-1}$ as in (\ref{cutoff}). Then the 
 asymptotic spot size of the resulting function
 is $O(\delta)$. In other words,
 the mixed mode TR of  near field recovers qualitatively 
 the asymptotic behavior of the Green function 
 near the source point.

When the superresolution terms from $K_{\rm DM}$ and $K_{\rm MD}$ combine in 
$K_{\rm PB}$, they cancel
each other resulting in diffraction-limited focal spot. 

In Figure \ref{fig:3d} the normalized transverse and longitudinal
profiles  are shown and
their resolutions  calculated for various TR schemes. The profiles
are normalized to be unity at the source location.  
 In the simulation, we use
$\lambda=2\pi$, $A=2\lambda$ and $\br_0=(0.01,0,0)$. 
Before normalization, the $K_{\rm M}$ curve
has the maximum less than $0.5$ and hence
is far from the square-root asymptotic regime discussed
 in the paragraph following (\ref{41-2}). 
 
 {
\subsection{Two-dimensional case}
 \begin{figure}
\begin{center}
\includegraphics[width=8cm, totalheight=6cm]{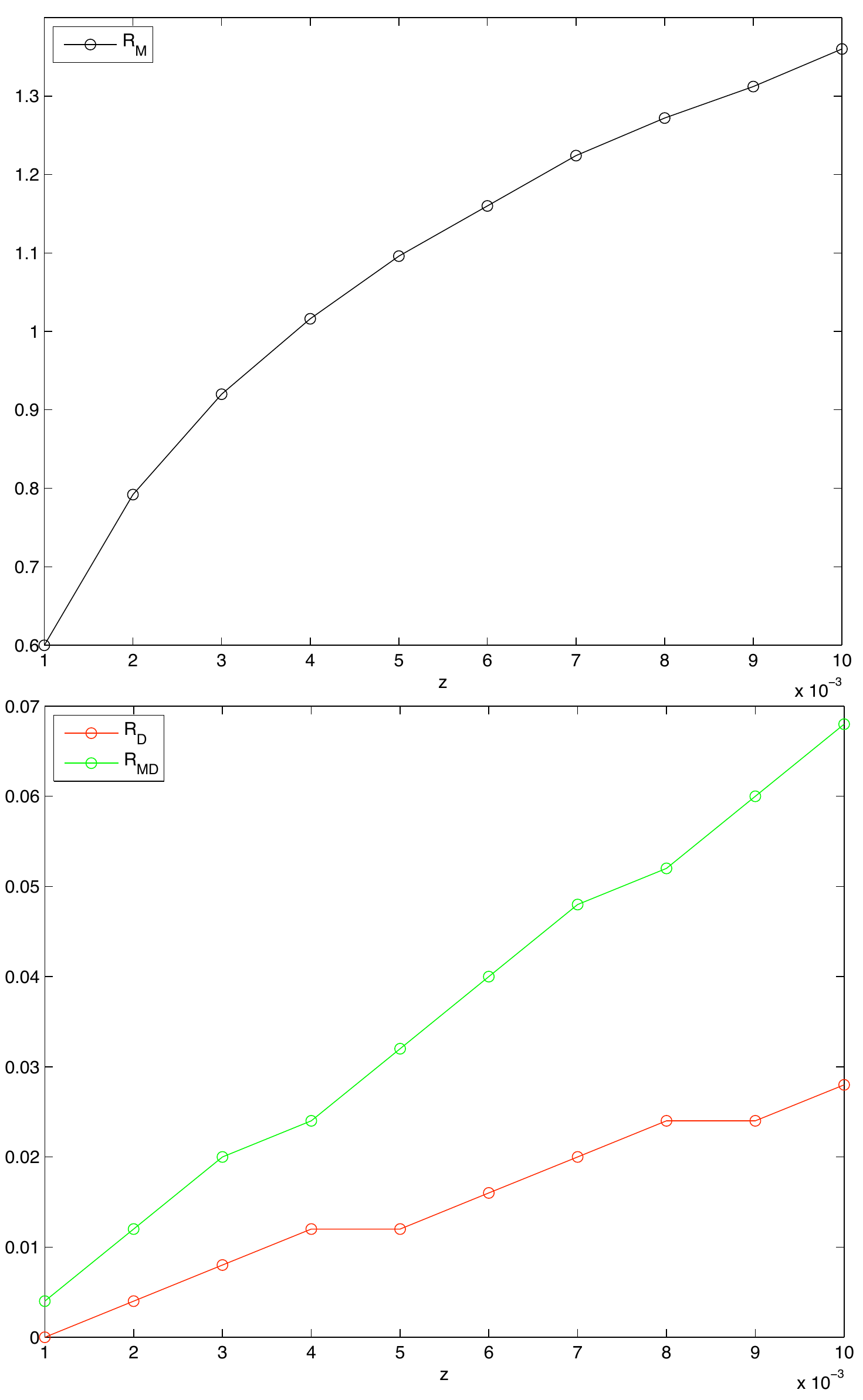}
\includegraphics[width=8cm, totalheight=6.1cm]{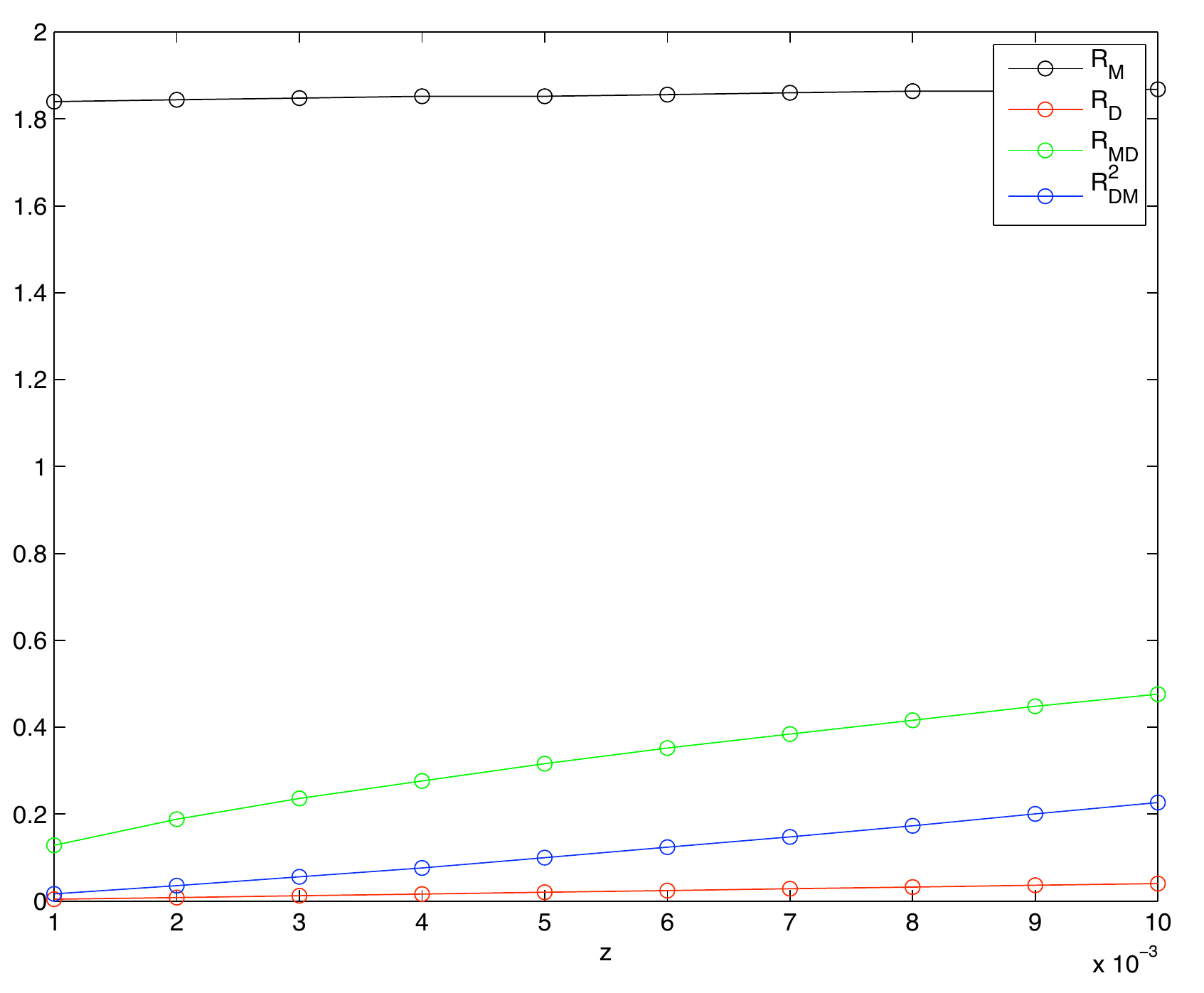}
\end{center}
\caption{The transverse resolution of various
TR schemes in three (left) and  two (right) dimensions. The linear behavior 
of $R^2_{\rm MD}$ in the right plot
 demonstrates the square-root
behavior of $R_{\rm MD}$ in two dimensions.  
 }
\label{fig:2d-res}
 \end{figure}

It is instructive to compare
the scaling behavior of resolution in two and three
dimensions.  In two dimensions, (\ref{24}) becomes
\beq
\label{24'}
{1\over k}\int {\delta\over x'^2+\delta^2}\times
{\delta\over (x-x')^2+\delta^2}dx'.
\eeq
The same calculation shows that 
\[
K_{\rm D}(\br,\br_0)\sim \delta^{1-2\alpha},\quad
\alpha\in [0,1] 
\]
and  $K_{\rm D}(\br_0,\br_0)\sim \delta^{-1}$. 
The $K_{\rm D}$-resolution asymptotic is given by $\delta^\alpha, \alpha=1$. 

For the mixed mode, (\ref{26}) and (\ref{27}) become
\beq
\label{26'}
&&{1\over k}\int \log{ (x'^2+\delta^2)^{-1/2}}\times
{\delta\over (x-x')^2+\delta^2}dx'\\
&& 
\label{27'}
-{1\over k}\int {\delta\over x'^2+\delta^2}\times
\log{((x-x')^2+\delta^2)^{-1/2}}dx'
\eeq
yielding 
\[
K_{\rm MD}(\br,\br_0), K_{\rm DM}(\br,\br_0)\sim \alpha 
\lt(-\log{\delta}\rt)^2,\quad
\alpha\in [0,1] 
\]
and  $K_{\rm MD}(\br_0,\br_0), K_{\rm DM}(\br_0,\br_0)\sim \lt(-\log{\delta}\rt)^2$. 
Hence, the  asymptotic  resolution for the mixed mode TR is given by $\delta^\alpha, \alpha=1/2$. Empirically, the square-root regime for the mixed mode TR in two dimensions
sets in much earlier than the monopole TR in three dimensions.  
Here lies the main difference
between the two and three dimensional cases. Namely, 
in two dimensions the
transverse  resolution for the mixed mode TR is proportional to the {\em square-root}
of the distance between the source and TRM. 

For the monopole TR in two dimensions, 
$K_{\rm M}$ is uniformly bounded 
w.r.t. $\delta$ and the resolution can not be deduced from
the asymptotic analysis. It is likely though that
the monopole TR resolution is qualitatively independent of $\delta$. 
This is amply confirmed by our numerical simulations (Figure \ref{fig:2d-res}). 

In Figure \ref{fig:2d-res} the transverse resolutions, denoted by
$R_{\rm X}, {\rm X=M,D, MD, DM}$, for the respective TR schemes  as a function
of the distance between the source and TRM are plotted
for 
the range $[0.001, 0.01]$. 
The dipole  resolution $R_{\rm D}$  exhibits  the linear
behavior  in two and three dimensions 
and the mixed mode resolution  $R_{\rm MD}, R_{\rm DM}$
exhibits the linear behavior in three dimensions
and square-root behavior in two dimensions. Both the dipole
and mixed mode resolutions  are  order
of magnitude better than the monopole resolution. 

In Figure \ref{fig:2d} the normalized transverse and longitudinal profiles
in two dimensions are shown and FWHM calculated for
various TR schemes with the same physical parameters
as in Figure \ref{fig:3d}. The unnormalized transverse profile
for the mixed mode TR has the maximum about 6.
  The dipole  resolution is much  better than the mixed mode   
  resolution which in turn is much better than the 
  monopole resolution
in both the transverse and longitudinal directions. 

 \begin{figure}
\begin{center}
\includegraphics[width=8cm, totalheight=6cm]{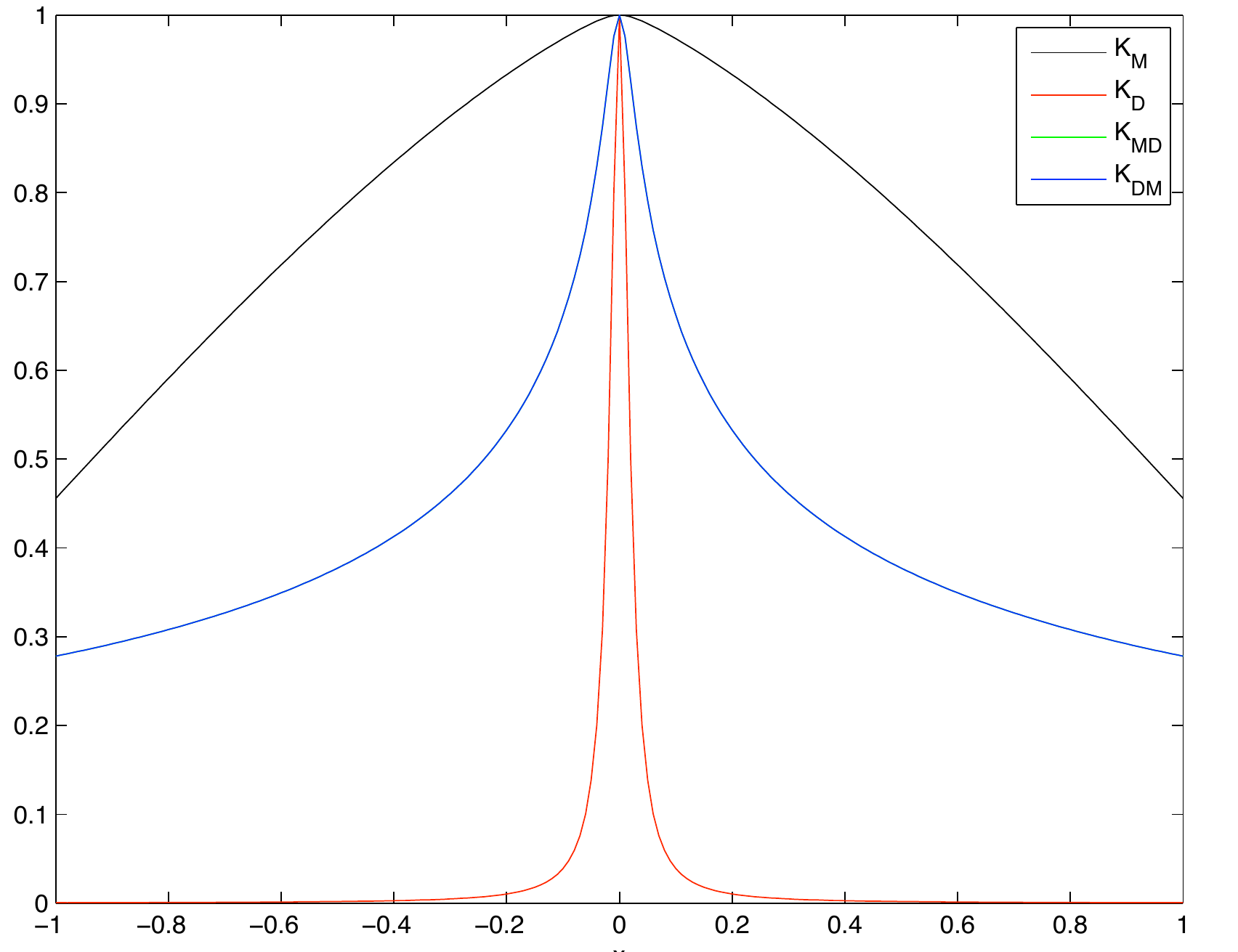}
\includegraphics[width=8cm, totalheight=6cm]{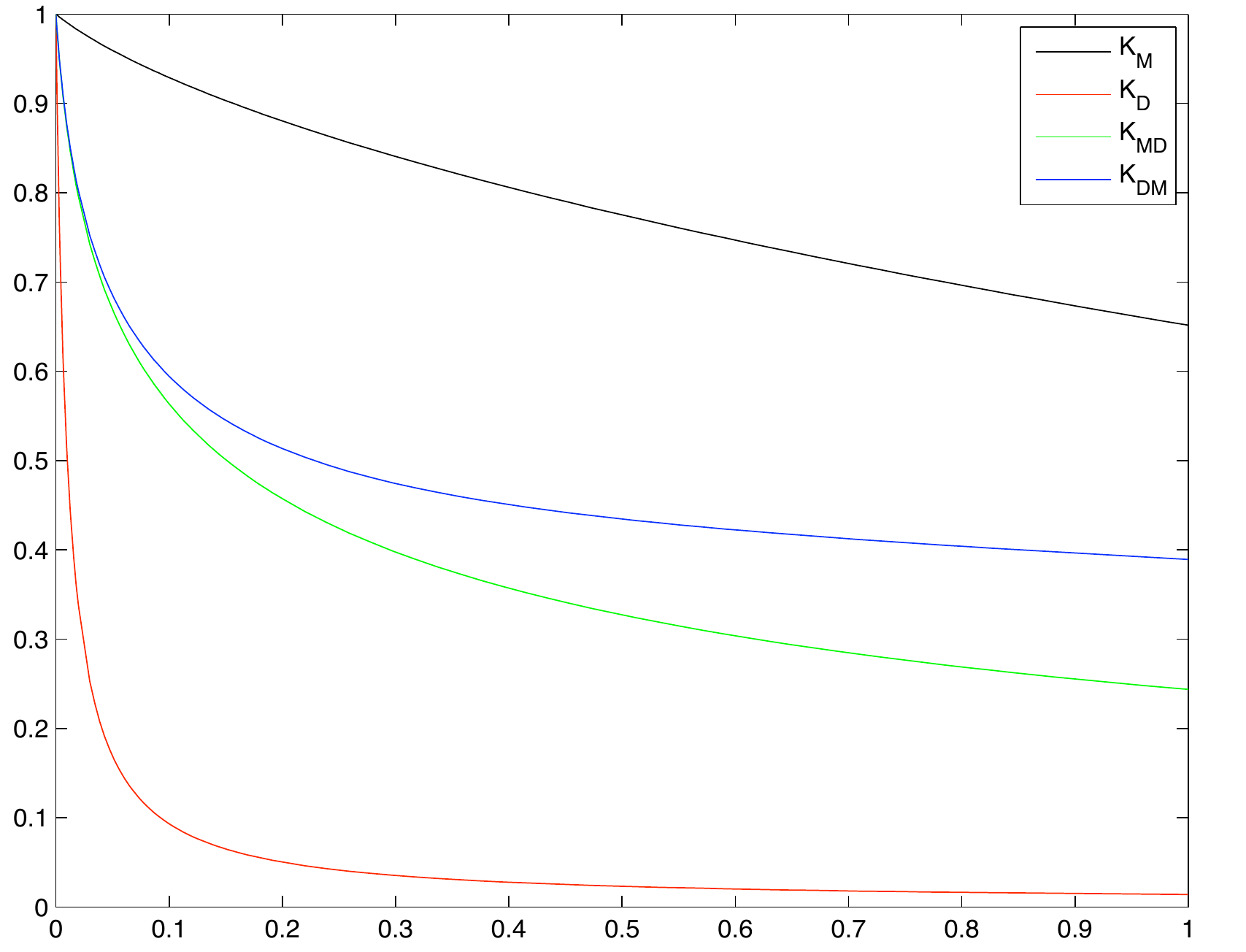}
\end{center}
\caption{(Left) 2d transverse  profiles with FWHM$= 1.87(K_{\rm M}), 0.05 (K_{\rm D}), 0.47 (K_{\rm DM}, K_{\rm MD})$; (Right)
2d longitudinal profiles: FWHM$=1 (K_{\rm M}), 0.009 (K_{\rm D}), 0.148 (K_{\rm MD}), 0.22 (K_{\rm DM}).$ Note that
the transverse profiles for 
the $K_{\rm MD}$ and $K_{\rm DM}$ coincide. 
 }
\label{fig:2d}
 \end{figure}

In summary, in the case of a monopole source the dipole TR produces  the best resolution
in both transverse and longitudinal directions, roughly
in  proportion to the distance between the source and TRM. 
The mixed mode TR can deliver nearly the same performance
in three dimensions and roughly the square-root resolution in the transverse direction  
in two dimensions. Between the two mixed mode TR schemes,
$K_{\rm MD}$ is still better than $K_{\rm DM}$ in
the longitudinal direction. 
A simple explanation for the superior
resolution of TRM involving the dipole field is that
a dipole source has a stronger singularity than a monopole source  (cf. (\ref{near3}) versus (\ref{near2}))
}

 \commentout{
 The main assumption for this conclusion is, of course,
 that the source is point-like or has a significantly subwavelength
 size. When the source has a size comparable to wavelength,
  $\Re[G]$ and $\Im[G]$ are 
 comparable and TR ensures that their effects add up
 rather than cancel out. 
 }

\section{Approximate TR is biased }
\label{sec4}

We have already seen that the perfect TR
described by 
(\ref{PB}) with $G=G_0$  is diffraction-limited
and contains no evanescent component of the initial outgoing field (see the remark following (\ref{weyl})). 

For the half space (domain (iii) of Figure \ref{domain}),
the result is similar: Suppose the initial wave field is
perfectly phase-conjugated on the TRM (i.e.
the boundary condition (\ref{BC})  is satisfied exactly by, say, 
the scheme $K_{\rm MD}$). 
Then in the interior of the half space the homogenous components of
 the initial wave are perfectly phase-conjugated but
 the evanescent components are exponentially
 damped by the factor
 \[
 e^{-2kz\sqrt{s_1^2+s_2^2-1}},\quad s_1^2+s_2^2>1
 \]
 depending on the distance $z$ from the TRM to
 the field point $(z,\bx)$ (Appendix \ref{sec:be}).
 
  On the other hand, for the paraxial wave (\ref{para}),
 perfect phase conjugation on one plane implies
 perfect phase conjugation in the half space
 due to the fact that the paraxial wave field is uniquely
 determined by the direction of propagation and
 the initial condition on one plane (Appendix \ref{sec:be}). 
 As a consequence, $v_{\rm M}$ and $ K_{\rm M}$
 are perfect TR schemes 
 in their respective contexts. In particular, 
 $K_{\rm M}$  can focus
 perfectly back on  the location of the point source
 in the paraxial approximation.

The perfect focusing of time reversed paraxial wave is, of course,
unphysical because the evanescent waves are not accounted
for. The other aspect that is more impractical than
unphysical with the  perfect conjugation of
the homogeneous component of the spherical wave,  is that an infinite aperture ($A=\IR^2$) is required for such a result.

In this section, we show 
that  when $D$ is a half space and $A$ is bounded, the maximum point of the
TR focal spot
does {\em not}
correspond to the exact source location $\br_0$ in general. 
In contrast, when $A=\IR^2$ in the paraxial regime, since  $K_{\rm M}$ 
is a perfect TR scheme for 
the paraxial wave,  the focal spot
is always centered. 

This bias effect tends to be much more pronounced in
the longitudinal direction. 
As the aperture of TRM increases, the transverse offset
(bias) 
decreases 
while the longitudinal offset may persist if there is
insufficient scattering in the medium, cf. Section \ref{sec6}. 
This is why we consider only the transverse resolution
in the previous section.

Let us recast the problem in  more general terms.  
Let $\phi(\br, \br')$ be  the phase of
  $G(\br,\br')$ and define
\beq
\label{ams}
P_{\rm M}(\br)
&=&\lt|\int_{A} G^*(\br',\br_0) e^{i\phi(\br,\br')}
W(\br, \br') d\sigma(\br')\rt|^2
  \eeq
  where $W(\br, \br')$ is a non-negative function.  The monopole TR kernel $K_{\rm M}$ 
  corresponds to $W(\br, \br')=|G(\br,\br')|$.
  
Direct calculation of $\nabla P_{\rm M}(\br_0)$ yields
\beqn
\lefteqn{\nabla P_{\rm M}(\br_0)}\\
&=&\int_{A\times A} |G^*(\br',\br_0)||G(\br'',\br_0)|
 W(\br_0, \br') 
 W(\br_0, \br'') (i\nabla \phi(\br_0,\br')-i\nabla \phi(\br_0,\br''))d\sigma(\br')d\sigma(\br'')\\
 &&+\int_{A\times A} |G^*(\br',\br_0)||G(\br'',\br_0)|
\nabla\lt( W(\br, \br') 
 W(\br, \br'')\rt)_{\br=\br_0} d\sigma(\br')d\sigma(\br'').
 \eeqn
 The first term vanishes due to anti-symmetry with respect to $\br',\br''$. 
 
 Consider the  case
 of a  planar TRM on the plane $z=0$
with  the paraxial wave  propagating in the
$z$-direction (Appendix A).
Suppose the medium inhomogeneities
  act like a phase object
affecting  only the phase of the free propagator. Namely,
$G(z,\bx,z',\bx')$ takes the form
\beq
\label{phase-obj}
G(z,\bx,z',\bx')=-
{1\over 4\pi (z-z')} e^{ik(z-z')} e^{i{k|\bx-\bx'|^2\over 2(z-z')}}e^{i\theta
(z,\bx,z',\bx')},\quad\bx,\bx'\in \IR^2.
\eeq
Note that for (\ref{phase-obj})  the expression (\ref{ams}) is divergent and thus
invalid at $\br=\br_0$ if $A=\IR^2$. 

For a finite aperture, $W(\br,\br')=(4\pi |z-z'|)^{-1}$
and thus  $\nabla\lt( W(\br, \br') 
 W(\br, \br'')\rt)_{\br=\br_0}$ points in the negative $z$ direction and consequently $\br_0$ tends to be farther away from $A$
 than the maximum point of $P_{\rm M}$. In other words, the
 focal spot of the monopole TR tends to be closer to $A$ than
 the source location. 
 Even in the case
 of a closed cavity 
 the offset (bias)  is reduced but not completely eliminated
 except for special cases. 

Now let us change the perspective and
think of TR as a means of imaging 
a  point source in a perfectly known medium. 
In this context a few widely used  $W(\br,\br')$'s
are:
\beqn
W(\br,\br')&=&|G(\br,\br')|^{-1}\quad \mbox{(the inverse filter)},\\
W(\br,\br')&=&W(\br')/\sqrt{\int_A |W(\br'')|^2d\sigma(\br'')},\,\,\forall W\in L^2(A, d\sigma)
\quad\mbox{(the phase processor)},\\
W(\br,\br')&=&|G(\br,\br')|/\sqrt{\int_A |G(\br,\br'')|^2d\sigma(\br'')}\quad\mbox{ (the conventional matched field processor)}.
\eeqn
For the inverse filter, $\nabla (W(\br,\br')W(\br,\br''))_{\br=\br_0}$ and thus $\nabla P_{\rm M}(\br_0)$  tend to
point away from the TR mirror $A$ and hence the source
location tends to be closer to $A$ than the maximum
point of $P_{\rm M}$. Again, when $A=\partial D$ or
in the far field regime this
offset can be reduced. 

Direct substitution shows that $\nabla P_{\rm M}(\br_0)=0
$ for the two  latter choices of $W$. Moreover,
  $\br_0$ is the maximum point of $P_{\rm M}$ for both
  choices of $W$. 
In the case of the $W$-weighted phase processor, we clearly have
\[
P_{\rm M}(\br)\leq \int_{A}| G^*(\br',\br_0)|
W(\br')d\sigma(\br')= P_{\rm M}(\br_0).
\]
In the case of the conventional matched field processing,
it is well known \cite{BKM, Tol} that
\beq
\label{mfp}
V(\br_0, \br')=G(\br_0,\br')/\sqrt{\int_A |G(\br_0,\br')|^2d\sigma(\br')}
\eeq
is the solution to the optimization problem: Maximize the quantity
\beqn
 \lt|\int_{A} G^*(\br',\br_0)
 V(\br') d\sigma(\br')\rt|^2
\eeqn 
over $V$ subject to the constraint 
\beq
\label{const}
\int |V|^2(\br')d\sigma(\br')=1.
\eeq 
Therefore
\[
P_{\rm M}(\br)=\int_A G^*(\br',\br_0) V(\br,\br')d\sigma(\br')
\]
is maximized at $\br=\br_0$ with the conventional
matched field processing.

In summary, as an imaging method both TR and the inverse filter are  generally biased; the former 
tends to underestimate the range of the  source  
while the latter tends to overestimate it.
The conventional matched field processor is unbiased and produces  the optimal
signal-to-noise ratio. The phase processor, which uses only
the phase of the Green function,  is a simple alternative
to the conventional matched field processor. 

Moreover,
in the regime described by (\ref{phase-obj}), 
$|G(\br,\br')|$ is a function of  the  distance between
the source point and the TRM.
In such a case the conventional matched field processor
coincides with  the phase processor with $W(\br')=\II_A/|A|^{1/2}$ 
independent of $\br'$.

 Similar conclusions can be drawn for the pure dipole case
with
\beq
\label{ams-dipole}
P_{\rm D}(\br)
&=&\lt|\int_{A} {\partial G^*(\br',\br_0)\over \partial n}
 e^{i\psi(\br,\br')}
W(\br, \br') d\sigma(\br')\rt|^2
  \eeq
  where $\psi(\br,\br')$ is the phase of 
  $\partial G (\br,\br')/\partial n$. $P_{\rm D}$ can
  be analyzed as above with $G$ replaced by its
  normal derivative.

For $K_{\rm MD}, K_{\rm DM}$, 
 consider the paraxial regime for which
(\ref{38}) is valid. 
 Each term on the right hand side of (\ref{38})
can be analyzed as before.
The result is that the mixed mode TR's are  biased, especially in 
 the longitudinal direction, just like  
 the monopole and dipole TR's.

\section{Randomness eliminates bias}
\label{sec6}
\begin{figure}
\begin{center}
\includegraphics[width=7.5cm, totalheight=5.7cm]{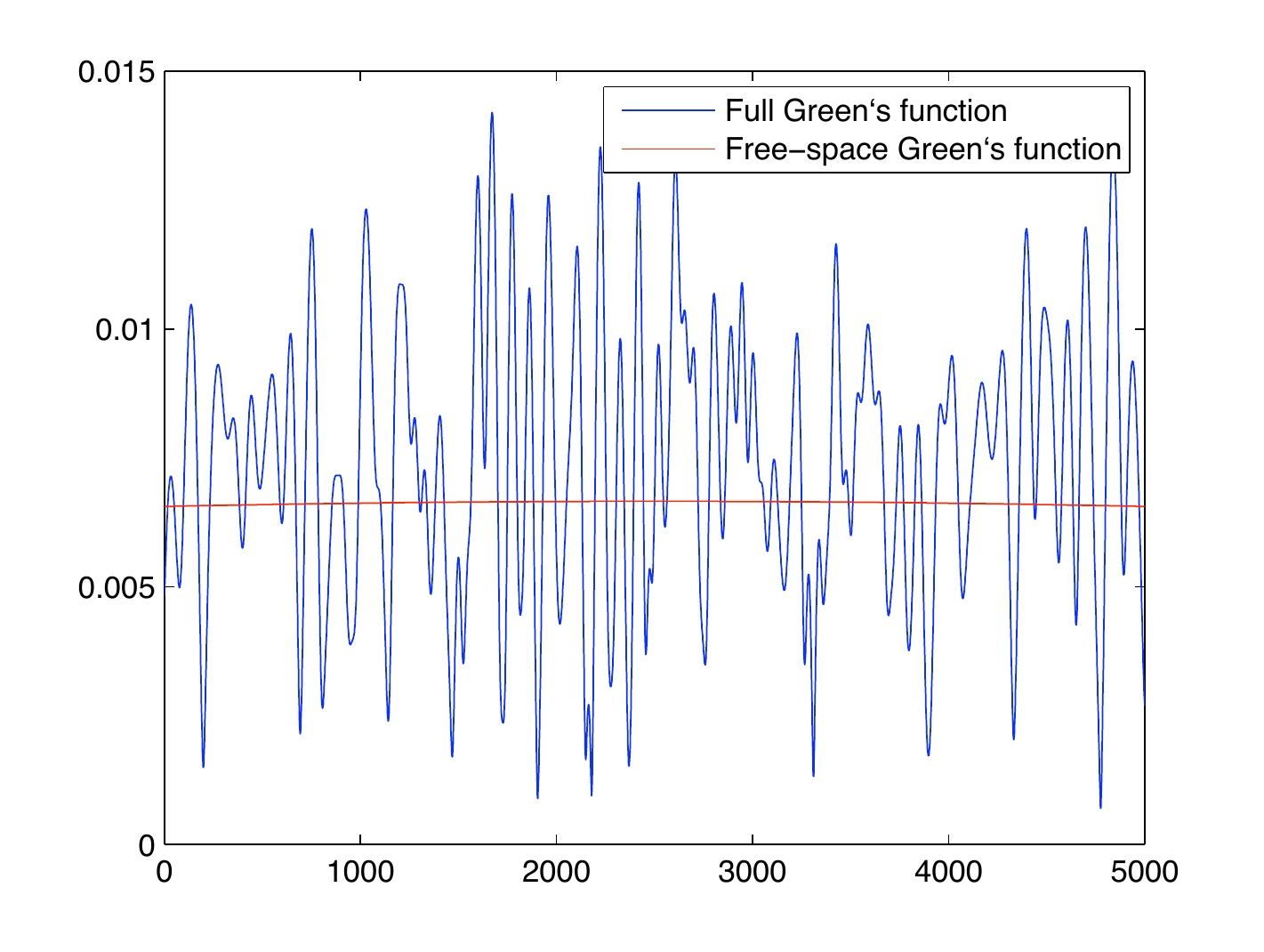}
\includegraphics[width=8.5cm, totalheight=6cm]{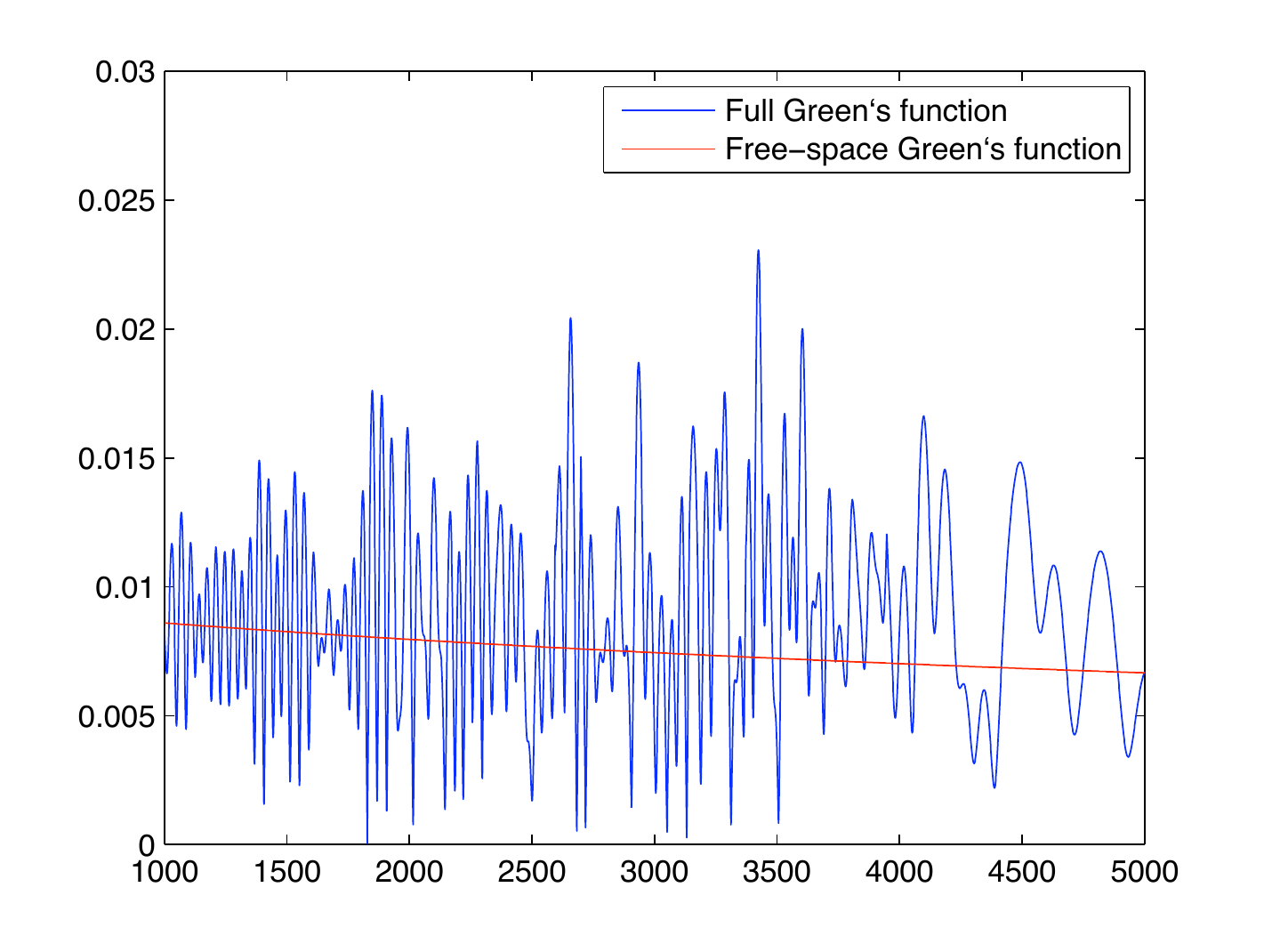}
\end{center}
\caption{The transverse (left), at $x=5000$,  and longitudinal (right), at $y=2500$,   profiles  of the intensity of
the $2-d$ Green function at wavelength $70$ with 1000 particles  randomly distributed in $[2000, 4000]\times [0,5000]$ and source (not shown) at
$(-5000, 2500)$
whose $n^2-1$
is $70$. (Reproduced from \cite{SA-rice}) }
\label{green-graph}
 \end{figure}

In this section, we explain  how randomness can make
the TR focal spot centered at the location of the source point.

Consider typical profiles
of the Green function in a multiple scattering medium as depicted
in Figure \ref{green-graph}.
A main feature of the intensity profiles is the rapid fluctuations
in the transverse and longitudinal directions. These fluctuations
upon differentiation and averaging (integration over $A\times A$) as in the expression
\beq
{\nabla P_{\rm M}(\br_0)}=
{1\over 2}\int_{A\times A} 
\nabla\lt|G(\br, \br') 
 G(\br, \br'')\rt|^2_{\br=\br_0} d\sigma(\br')d\sigma(\br'')
 \label{rand}
 \eeq
 essentially cancel one another. 
 
 More precisely, the degrees of scattering and randomness
 can be measured by the spatial spread of the propagation channel. 
 If the spatial spread increases
 to infinity, then the TRM acquires an {\em effective} aperture
 that is proportional to the spatial spread \cite{BPZ, tire-dual}. Then, as per discussion at the beginning of Section \ref{sec4},
 the accuracy of approximate TR's  improve and
 the focal spot becomes more centered and sharper.

Moreover, counterintuitively,  the TR image can avoid statistical fluctuations
if the medium is sufficiently random. Let us recall
some analytical  results from the literature.  

Consider TR 
in a random medium occupying a half space 
 with TRM on the planar boundary. 
When the random channel is the Rayleigh
fading regime (i.e. obeying zero-mean Gaussian statistics) with a divergent spatial spread
the monopole TR (\ref{LB})  with TRM's aperture much larger
than the coherence length $\ell_c$ will focus on the source point with
a spot size $\sim \ell_c$ \cite{tire-hole, pulsa}. 
The conventional coherence theory suggests that the
coherence length is of the order of wavelength \cite{She}.
However, spatial correlations of much smaller
extent has been recently measured in the near field
of random media \cite{AD}. This effect in conjunction with
the preceding theory  may be the key to
understanding  the subwavelength focusing observed
in the time reversal experiment reported in \cite{LRF}.

 \commentout{
 What eventually  eliminates the bias  in the monopole TR is the divergence of
 the spatial spread which results in a vanishing
   $\nabla\lt( W(\br, \br') 
 W(\br, \br'')\rt)_{\br=\br_0}$ in the expression for $\nabla
 P_{M}(\br_0)$. 
 }
 
The  result of stability and focusing
is a special case of what was 
originally established for   the setting of 
multiple frequencies and multiple source points \cite{tire-hole,pulsa}.
With straightforward modification of the arguments in \cite{tire-hole, pulsa} the same result  can be extended to 
any TR of  the types
 (\ref{LB})-(\ref{DM}) in the paraxial regime, the only difference 
 being in  the asymptotic shape of the focal spot.  
Experiments 
show consistently similar results on stability and focusing
properties \cite{Fin, LRF,  YPY}.  
 
 \commentout{
 For the spherical wave in multiple scattering media 
it is generally accepted that the coherence length $\ell_c$
is on the order of $\lambda$. If so, then a  subwavelength focusing
can not be realized with approximate TR on  {\em one} plane.
}

 \commentout{
 On the contrary, neither stability nor focusing is attainable
 for the ``perfect'' TR represented by (\ref{PB}) 
 since in the Rayleigh fading regime the Green function
 has a zero mean implying that  $\IE K_{\rm PB}=0$. 
 }
 
 \commentout{
 Within the scope of  paraxial wave (\ref{para})  under suitable
 conditions \cite{BPZ} 
 the (transverse) coherence length $\ell_c$  can 
 be much smaller than the wavelength, indeed 
 inversely proportional to the spatial spread
 as the latter diverges \cite{tire-pla}. 
This, however, may be  an artifact of the paraxial wave 
as approximation of the spherical wave
since in the simplest case of $ n=1$ the classical Rayleigh resolution formula 
($\lambda L/A$
 where $A$ is the TRM's aperture and $L$ the distance of
the source point to TRM) already predicts an arbitrary subwavelength resolution (e.g. as $A\to\infty$) which
is never observed in real experiments.  

}

In conclusion, random fluctuations in the intensity
of Green's function is a necessary condition  for
the desired focusing property of time reversal with
a finite aperture.

\section{Conclusion}
We have reexamined various time reversal schemes that
can be employed from the boundary of a domain; they involve
monopole and/or dipole fields.

We have
explored their relationships  in the paraxial regime. For the boundary
 behavior,  we show that
 for paraxial wave only the monopole TR scheme satisfies the
exact boundary condition while for the spherical wave
only the MD-mode TR scheme satisfies the
exact boundary condition. 
We have also
 shown
that  for the paraxial wave the standard  monopole TRM
produces the perfect result in the entire domain but
for the near field the monopole TRM can hardly achieve
subwavelength focusing. On the other hand, the dipole TRM is
 approximate in the paraxial regime
but is capable of producing  a focal spot size linearly proportional to
the 
distance from TRM to the source location, thus
focusing on a subwavelength scale  when the TRM is sufficiently close to  the source point.  The mixed mode TRM has  the similar (linear)
behavior  in three dimensions but the square-root asymptotic  in two dimensions.  The monopole TRM is, if possible at all, 
ineffective to focus  below wavelength.

We have seen how
two mixed mode effects combine in the so called
perfect TR scheme for a closed cavity and cancel
their respective subwavelength focusing property. 
The perfect TRM is ``perfect'' only for a source-free initial
field.

We have examined  an often neglected effect
of biased focusing associated with TRM of a finite aperture
in a weak- or non-scattering medium. This effect pertains
to  all five TR schemes discussed in the paper.
In a multiply scattering medium, the bias is greatly
reduced and the focal spot becomes centered at the source location. The removal of bias by random scattering
is attributed to the enlarged effective aperture and
the random fluctuations in the {\em intensity} of the Green
function.

 \begin{appendix}
 \section{Mixed mode TR for paraxial wave}

Let $z\in \IR$ be the longitudinal coordinate and $\bx\in \IR^2$
the transverse coordinates. 
Let $G^\pm_{\rm P}(z,\bx,z',\bx')$ be the Green functions for 
the paraxial wave equation 
\beq
\label{para}
\pm i\partz \Psi+\frac{1}{2k}\Delta_{\rm T}\Psi+{k\over 2}(n^2(z,\bx)-1)\Psi=0
\eeq
where 
$\Delta_{\rm T}$ is the transverse Laplacian.
The plus sign represents positive $z$ propagating wave
and minus sign negative $z$ propagating wave.  

\commentout{
Since $G^\pm_{\rm P}$ is the Green function for (\ref{para}), 
 $\partial G^\pm_{\rm P}/\partial z'$ satisfies the same equation
 with the initial data
 \beq
 \label{del}
 {\partial 
 \over\partial z'} G^\pm_{\rm P}(z,\bx,z',\bx')|_{z=z'}
 &=&\mp i\lt[{1\over 2k}\Delta_{\rm T} +{k\over 2}(n^2(z',\bx)-1)\rt]
 \delta(\bx-\bx').
 \commentout{
  {\partial 
 \over\partial z} G^-_{\rm P}(z,\bx,z',\bx')|_{z=z'}
 &=& i\lt[{1\over 2k}\Delta_{\rm T} +k\tilde n(z',\bx)\rt]
 \delta(\bx-\bx')
 }
 \eeq
By  (\ref{del}) and  integrating by parts we can also show
that  $G^\pm_{\rm P}(z,\bx,z',\bx')$ as a function of $z',\bx'$ satisfies the same equation as (\ref{para}). 
}

The full Green function $G(z,\bx,z',\bx')$ 
in the paraxial regime can be expressed as 
\beq
\label{phase2}
G(z,\bx,z',\bx')& =&{ie^{ik(z-z')}\over 2k} G^-_{\rm P}(z, \bx,z',\bx'),\quad z>z'\\
G(z,\bx,z',\bx')& =&{ie^{ik(z'-z)}\over 2k} G^-_{\rm P}(z,\bx,z',\bx'),\quad z<z'
\eeq

Let $D=\{z>0\}$, the plane  $z=0$ be occupied by the TRM and $\br_0=(z_0,\bx_0)$ be the location of
the source. 
 Assume that the TRM is away from the medium inhomogeneities
  i.e. $ n(0,\bx)=1$.
 Then $K_{\rm MD}$ can
 be written as
 \beq
 \label{a1}
 K_{\rm MD}(\br,\br_0)&=&-{ie^{ik(z-z_0)}\over 4k}\int G_{\rm P}^{-*}(0,\bx', z_0,\bx_0) G^-_{\rm P}(0, \bx',  z,\bx)d\bx'\\
 &&+{e^{ik(z-z_0)}\over 4k^2} \int G_{\rm P}^{-*}(0,\bx', z_0,\bx_0) {\partial 
 \over\partial z'} G^-_{\rm P}(z',\bx',z,\bx)|_{z'=0}d\bx'.\nn
 \eeq
  By the assumption $n(0,\bx)=1$ and (\ref{para}),
 we have 
 \beq
 K_{\rm MD}(\br,\br_0)&=&-{ie^{ik(z-z_0)}\over 4k}\int G_{\rm P}^{-*}(0,\bx', z_0,\bx_0) G^-_{\rm P}(0,\bx', z,\bx)d\bx'\\
 &&+{ie^{ik(z-z_0)}\over 8k^3} \int G_{\rm P}^{-*}(0,\bx', z_0,\bx_0) \Delta'_{\rm T} G^-_{\rm P}(0,\bx', z,\bx)d\bx'\nn\\
 &=&-ik\int G^{*}(0,\bx', z_0,\bx_0) G(0,\bx', z,\bx)d\bx'\nn\\
 &&+{i\over 2k} \int \nabla'_{\rm T} G^{*}(0,\bx', z_0,\bx_0)\cdot \nabla'_{\rm T} G(0,\bx', z,\bx)d\bx'\nn
 \eeq
 after integration by parts. The result for
 $K_{\rm DM}$ 
 can be likewise derived. 
 \section{Boundary values}
 \label{app:bc}
 \subsection{Paraxial wave}
 Let us first focus on the boundary values of
 (\ref{LB'})-(\ref{DM'}). Let $u(z,\bx)$ be
 a paraxial wave propagating in the negative $z$ direction
 which can be written as 
 \[
 e^{-ikz} \Psi(z,\bx)
 \]
 where $\Psi$ is the solution of eq. (\ref{para}) with
 the minus sign.

 In the paraxial regime with $ n(0,\bx)=1$, we have
 from (\ref{phase2}) and (\ref{para}) that
 \beqn
  \label{LB''}
  v_{\rm M}(0,\bx)&=&
  \int u^*(0,\bx')G^+_{\rm P}(0,\bx,0,\bx')d\bx'\\
    \label{DP''}
  v_{\rm D}(0,\bx)&=&-{i\over k}
\int \Delta_{\rm T} u^*(0,\bx')G^+_{\rm P}(0,\bx,0,\bx')d\bx' 
-{i\over 2k^3} \int \Delta_{\rm T}^2 u^*(0,\bx')  G^+_{\rm P}(0,\bx,0,\bx')d\bx'
\\
\label{MD''}
v_{\rm MD}(0,\bx)&=&{1\over 2}
\int u^*(0,\bx')G^+_{\rm P}(0,\bx,0,\bx')d\bx' 
+{1\over 4k^2} \int\Delta_{\rm T} u^*(0,\bx')G^+_{\rm P}(0,\bx,0,\bx')d\bx'\\
v_{\rm DM}(0,\bx)&=&{1\over 2}
\int u^*(0,\bx')G^+_{\rm P}(0,\bx,0,\bx')d\bx' 
+{1\over 4k^2} \int\Delta_{\rm T} u^*(0,\bx')G^+_{\rm P}(0,\bx,0,\bx')d\bx'
\label{DM''}
  \eeqn
  where $u$ is a paraxial wave propagating
  in the negative $z$ direction. 
 Setting $G^\pm_{\rm P}(0,\bx,0,\bx')=\delta(\bx-\bx')$, we
 obtain 
 the following boundary conditions
\beqn
\label{BC-para1}
 v_{\rm M}(0,\bx)&=& u^*(0,\bx)\\
 v_{\rm D}(0,\bx)&=& -{i\over k} \Delta_{\rm T} u^*(0,\bx')-{i\over 2k^3} \Delta_{\rm T}^2 u^*(0,\bx')\label{BC-para2}\\
 v_{\rm MD}(0,\bx)&=&
 {1\over 2} u^*(0,\bx)+{1\over 4k^2} \Delta_{\rm T} u^*(0,\bx)\label{BC-para3}\\
 v_{\rm DM}(0,\bx)&=&{1\over 2} u^*(0,\bx)+{1\over 4k^2} \Delta_{\rm T} u^*(0,\bx)
\label{BC-para4}\\
 v_{\rm PB}(0,\bx)&=& u^*(0,\bx)+{1\over 2k^2} \Delta_{\rm T} u^*(0,\bx)\label{BC-para5}.
 \eeqn
 The kernel $K_{\rm X}, {\rm X}={\rm M,D,MD, DM, PB}$ satisfy
 the similar boundary properties.
 \commentout{
 \beqn
\label{BC-para1'}
 K_{\rm M}(0,\bx, z_0,\bx_0)&=& G^*(0,\bx, z_0,\bx_0)\\
 K_{\rm D}(0,\bx, z_0,\bx_0)&=& -{i\over k} \Delta_{\rm T} G^*(0,\bx,z_0,\bx_0)-{i\over 2k^3} \Delta_{\rm T}^2 G^*(0,\bx,z_0,\bx_0)\label{BC-para2'}\\
 K_{\rm MD}(0,\bx, z_0,\bx_0)&=&
 {1\over 2} G^*(0,\bx,z_0,\bx_0)+{1\over 4k^2} \Delta_{\rm T} G^*(0,\bx, z_0,\bx_0)\label{BC-para3'}\\
 K_{\rm DM}(0,\bx, z_0,\bx_0)&=&{1\over 2} G^*(0,\bx,z_0,\bx_0)+{1\over 4k^2} \Delta_{\rm T} G^*(0,\bx, z_0,\bx_0)
\label{BC-para4'}\\
 K_{\rm PB}(0,\bx, z_0,\bx_0) &=& G^*(0,\bx, z_0,\bx_0)+{1\over 2k^2} \Delta_{\rm T} G^*(0,\bx, z_0,\bx_0).\label{BC-para5'}
 \eeqn
 }
 In other words, only the monopole  TR
 gives
the correct boundary values in the case of
paraxial wave propagating in the half space.
 \commentout{
 \beqn
 v_{\rm M}(0,\bx)&=& u^*(0,\bx)\\
 v_{\rm D}(0,\bx)&=& -{i\over k} {\partial \over\partial z} u^*(0,\bx')+{1\over k^2} {\partial^2 \over\partial z^2} u^*(0,\bx')\\
 v_{\rm MD}(0,\bx)&=&
 {1\over 2} u^*(0,\bx)+{i\over 2k} {\partial \over \partial z}
 u^*(0,\bx)\\
 v_{\rm DM}(\br)&=&- {i\over 2k} {\partial \over \partial z}
 u^*(0,\bx). 
 \eeqn
 }

\subsection{Spherical wave}

Consider the half space  $D=\{z>0\}$
with $A=\{z=0\}$. Assume $G=G_0$, the free space Green function.

Differentiating the Weyl representation for $G_0$
with respect to $z'$,
we find that 
\beq
\label{a32}
{\partial \over \partial z'}G_0(z,\bx,z',\bx')\lt.\rt|_{z=z'=0}={1\over 2} \delta(\bx-\bx'). 
\eeq 
From (\ref{a32}) it is immediately clear that
\[
v_{\rm MD}(0,\bx)={1\over 2} u^*(0,\bx),\quad K_{\rm MD}(0,\bx)
={1\over 2} G^*(0,\bx, z_0,\bx_0). 
\]


However, neither $v_{\rm DM}$ nor $K_{\rm DM}$ satisfies
the boundary condition as we show now.

In the neighborhood of
$A$, the wave field $u$ in
$D$ can be represented in terms of the plane wave spectrum:
\beq
\label{plane-wave}
u(x,y,z) &=&\int {ds_1ds_2} 
 e^{ik(s_1x+s_2y)}
 e^{iks_3z}
A(s_1,s_2)
+\int {ds_1ds_2} 
 e^{ik(s_1x+s_2y)}
 e^{-iks_3z}
B(s_1,s_2)
\eeq
where $s_3$ is given  in (\ref{weyl}). 
Consider for simplicity one-way wave with, say, $B\equiv 0$.
Then the phase-conjugated field has the form
\beq
u^*(x,y,z) &=&\int_{s_1^2+s_2^2\leq 1} {ds_1ds_2} 
 e^{ik(s_1x+s_2y)}
 e^{-iks_3z}
A^*(-s_1,-s_2)\nn\\
&&+
\int_{s_1^2+s_2^2> 1} {ds_1ds_2} 
 e^{ik(s_1x+s_2y)}
 e^{iks_3z}
A^*(-s_1,-s_2)\nn
\eeq
and thus
\beqn
{\partial \over\partial z} u^*(x,y,0) &=&ik\int {ds_1ds_2} 
 e^{ik(s_1x+s_2y)}s_3
A^*(-s_1,-s_2)\lt[-I_{\{s_1^2+s_2^2\leq 1\}}+I_{\{s_1^2+s_2^2>1\}}\rt].\label{a20}
\commentout{
&&+\int_{s_1^2+s_2^2\leq 1} {ds_1ds_2} 
 e^{ik(s_1x+s_2y)}
 e^{iks_3z}
B^*(-s_1,-s_2)\nn\\
&&+
\int_{s_1^2+s_2^2> 1} {ds_1ds_2} 
 e^{ik(s_1x+s_2y)}
 e^{-iks_3z}
B^*(-s_1,-s_2).
} 
\eeqn
where $I_S$ is the indicator function of the set $S$. 
By Plancherel's identity we obtain
\beqn
v_{\rm DM}(0,\bx)&=&
2\pi \int {ds_1ds_2} 
 e^{ik(s_1x+s_2y)}
A^*(-s_1,-s_2)\lt[-I_{\{s_1^2+s_2^2\leq 1\}}+I_{\{s_1^2+s_2^2>1\}}\rt]\\
&\neq&2\pi u^*(0,\bx).
\eeqn
Namely, no constant multiple of $v_{\rm DM}$ satisfies
the boundary condition. 

Likewise, one can show that
none of the other TR schemes than $ v_{\rm MD}, K_{\rm MD}$ 
 satisfies  the boundary condition, with some adjustable constant factor. 

\commentout{
Next, we show that when any medium inhomogeneities
are present, i.e. $n^2\neq 1$, 
even $2\times v_{\rm MD}, 2\times K_{\rm MD}$ 
fail to satisfy the boundary condition.

 We assume
that the medium inhomogeneities $n^2-1$ is away
from the TRM so that the gap $D^-$ is a non-empty infinite slab.
Figure \ref{fig:half-space}. Then 
it can be shown  \cite{subwave-tr} that
the scattered field $u_s$ in $D^-$ can be represented
as
\[
u_s=
\int {ds_1ds_2} 
 e^{ik(s_1x+s_2y)}
 e^{-iks_3z}
B(s_1,s_2),\quad \br \in D^-
\]
for some angular spectral amplitude $B$. 
Hence the full Green function $G$ can be expressed
as
\beq
\label{a33}
G(\br,\br')&=&G_0(\br,\br')
+\int {ds_1ds_2} 
 e^{ik(s_1x+s_2y)}
 e^{-iks_3z}
B(s_1,s_2, \br'),\quad \br \in D^-,\quad  \br'\in \{z=0\}. 
\eeq
From (\ref{a32}) we obtain
\beq
\label{a34}
{\partial \over \partial z'}B(s_1,s_2, z'=0,\bx')=0,\quad\forall s_1, s_2, x', y' \in \IR
\eeq
if $G$ satisfies the same property (\ref{a32}). 
}

\commentout{
Consider the case $n=n(z)$. Then $G(\br,\br')=
G(z,\bx-\bx', z')$ must be translation-invariant in the transverse plane which implies
\beq
\label{a35}
B(s_1,s_2,\br')=B(s_1,s_2, z') e^{-ik(s_1x'+s_2y')}.
\eeq
By this and  (\ref{a34}) we find that $B\equiv 0$. 
}

\section{One-plane TRM in a half space}
 \label{sec:be}

\begin{figure}[t]
\begin{center}
\includegraphics[width=10cm, totalheight=6cm]{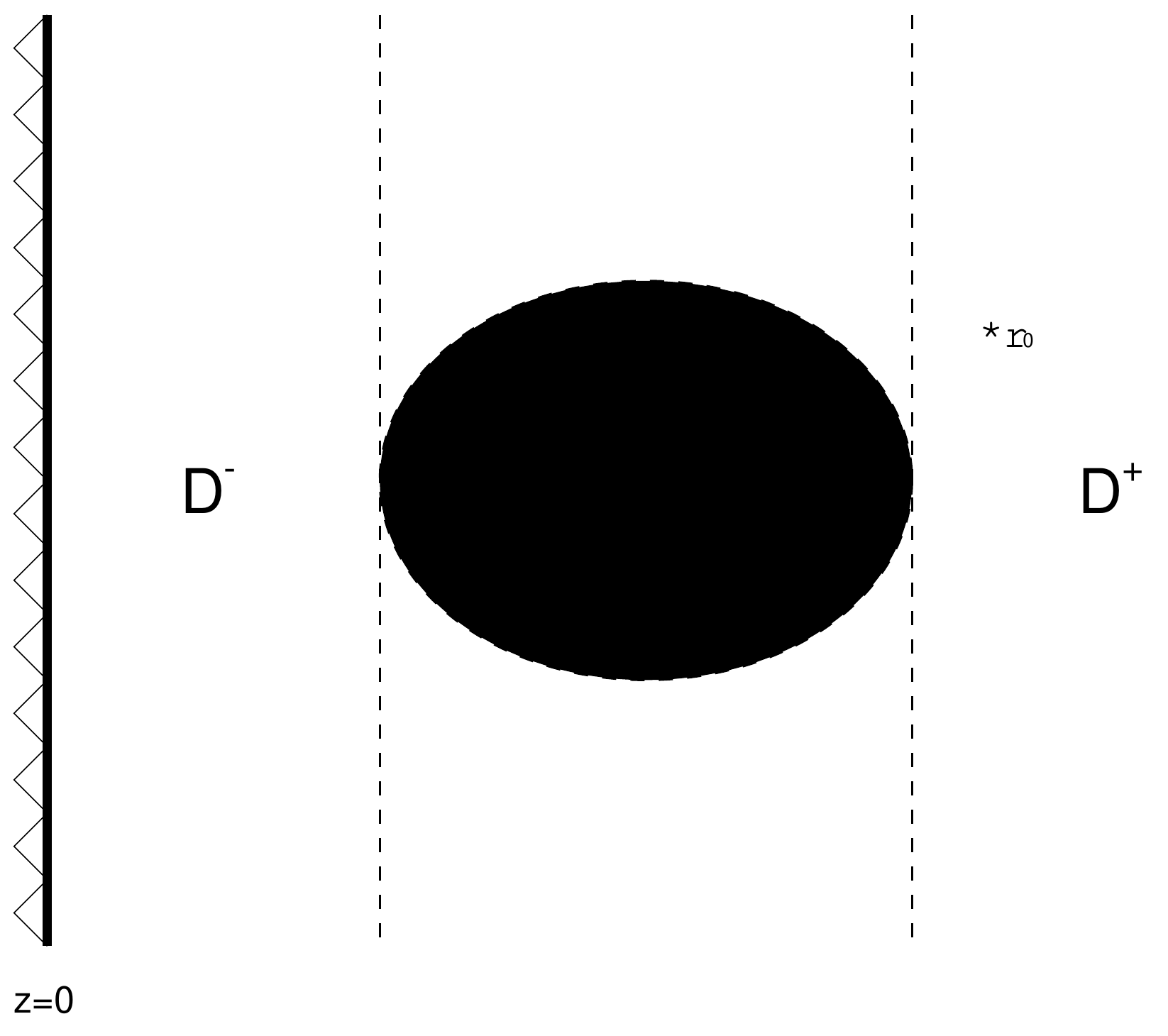}
\end{center}
\caption{The dark area represents the scattering domain $S$.
The asterisk, $*$,  represents the source point $\br_0$ which may be anywhere in $D$.}
\label{fig:half-space}
 \end{figure}

Consider the half space  $D=\{z>0\}$
with $A=\{z=0\}$ and point source located at $(z_0,\bx_0), z_0>0$, Figure \ref{fig:half-space}. 
 We assume
that the medium inhomogeneities $n^2-1$ is away
from the TRM so that the gap $D^-$ is a non-empty infinite slab.
Figure \ref{fig:half-space}. 

First we observe that for the paraxial wave, perfect phase conjugation
on $z=0$ implies perfect phase conjugation throughout 
the domain $D$. This can be shown as follows. 

Let $u(z,\bx)=e^{-ikz} \Psi^-(z,\bx)$ be the initial wave
propagating in the negative $z$ direction where $\Psi^-$
satisfies (\ref{para}) with the minus sign. If $u(0,\bx)$ is
perfectly conjugated by, say, either the scheme described
by $v_{\rm M}$ or
$v_{\rm PB}$, then the resulting field, denoted by $v(z,\bx)$,
is given by
\[
v(z,\bx)=e^{ikz}\Psi^+(z,\bx)
\]
where $\Psi^+$ satisfies (\ref{para}) with the plus sign
and the initial condition
\[
\Psi^+(0,\bx)=\Psi^{-*}(0,\bx).
\]
Now since $\Psi^{-*}(z,\bx)$ satisfies 
(\ref{para}) with the plus sign and
the initial condition $\Psi^+(0,\bx)$, we have from
the uniqueness theorem for the initial value problem
of (\ref{para}) that 
\[
\Psi^+(z,\bx)=\Psi^{-*}(z,\bx)
\]
and consequently
\[
v(z,\bx)=u^*(z,\bx),\quad z>0.
\]
The same argument and conclusion apply to
the paraxial wave with a point source. 

The case with the spherical wave is similar but
more subtle since the evanescent wave is involved. 
It can be shown  that 
the Green function
in $D^-\cap \{z<z_0\}$ can be expressed
as superposition of  plane waves propagating in the negative $z$ direction 
\beq
G(\br,\br_0)=
\int {ds_1ds_2} 
 e^{ik(s_1x+s_2y)}
 e^{-iks_3z}
B(s_1,s_2),\quad \br \in D^-\cap \{z<z_0\}
 \label{62}
 \eeq
 in analogy to (\ref{plane-wave}). 
 
Let  $W$  be a source-free field  propagating across $z=0$ into $D$ and  be represented as 
\beq
W(\br)=
\int {ds_1ds_2} 
 e^{ik(s_1x+s_2y)}
 e^{iks_3z}
A(s_1,s_2),\quad \br \in D^-.
 \label{61}
 \eeq
 Note the sign in the exponent associated with $z$ here is different
 from that in (\ref{62}) since they propagate in the opposite
 directions. 
 Suppose $W$ is the phase conjugate field of $G$ at $z=0$, i.e.
 $W^*(x,y,0)=G(x,y,0,\br_0), \forall x,y.$
 For $z=0$, equating (\ref{61}) with (\ref{62}) conjugated we obtain the condition 
 \beq
 A(s_1,s_2)&=& B^*(-s_1,-s_2).
 \label{evan-pc}
 \eeq
 This result is first observed in \cite{NW}. 

 Let $W_H$ and $W_E$ be, respectively,  the homogeneous and 
 the evanescent components of $W$, corresponding to, respectively, the integration (\ref{61}) restricted to
 $s_1^2+s_2^2\leq 1$ and $s_1^2+s_2^2> 1$. The homogeneous and the
 evanescent components $G_H, G_I$ of (\ref{62}) are
 defined analogously. 
 Then (\ref{evan-pc}) lead to, respectively,
 \beq
W_H^*(\br)&=&G_H(\br),\quad\br\in D^-\\
W_I^*(\br)&=&
\int {ds_1ds_2} 
 e^{ik(s_1x+s_2y)}
 e^{-kz\sqrt{s_1^2+s_2^2-1}}
B(s_1,s_2),\quad \br \in D^-.
 \eeq
 In other words, the homogenous components of
 the initial wave is perfectly phase-conjugated but
 the evanescent components are exponentially
 damped by the factor
 \[
 e^{-2kz\sqrt{s_1^2+s_2^2-1}}
 \]
 depending on the distance from the TRM to
 the field point.

 \bigskip
 
 {\bf Acknowledgement}. I thank my student Arcade Tseng
 for preparing Figures \ref{fig:3d}, \ref{fig:2d-res} and \ref{fig:2d}.
 
 \bigskip

  \end{appendix}

\end{document}